\documentclass[a4paper,12pt, epsfig]{article}
\def\letter{0}\def\pr{0}
\pdfoutput=1 
\usepackage{epsfig}
\usepackage{epstopdf}
\usepackage{textcomp}
\usepackage{graphicx}
\usepackage{ifthen}
\usepackage{ulem}

\usepackage{amsmath}
\usepackage[psamsfonts]{amssymb}
\usepackage{euscript}

\usepackage{latexsym}
\usepackage[arrow,matrix,curve]{xy}

\usepackage[hypertexnames=false]{hyperref}
\hypersetup{
    colorlinks=true,
    linkcolor=blue,
    filecolor=magenta,   
    citecolor=black,   
    urlcolor=cyan,
    }

\newskip\humongous \humongous=0pt plus 1000pt minus 1000pt

\newif\ifdtup

\def\,{\hspace{-.1cm}}
\def\hsp{,\hspace{.7cm}}

\def\fc#1#2 {\frac{n}{q}#1\frac{n}{q}#2}

\newcommand{\vac}{\ensuremath{|0\rangle}}

\renewcommand{\sin}{\textrm{sin}}

\renewcommand{\sinh}{\textrm{sinh}}

\renewcommand{\tanh}{\textrm{tanh}}
\newcommand{\sech}{\textrm{sech}}
\newcommand{\csch}{\textrm{csch}}

\renewcommand{\theequation}{\arabic{section}.\arabic{equation}}
\renewcommand{\(}{\begin{equation}}
\renewcommand{\)}{end{equation} \vspace{-.05in}\linebreak}

\newcounter{saveeqn}
\newcounter{savealpheqn}

\newcommand{\alpheqn}{\setcounter{saveeqn}{\value{equation}}%
  \stepcounter{saveeqn}\setcounter{equation}{0}%
  \renewcommand{\theequation}{\mbox{\arabic{section}.\arabic{saveeqn}
\alph{equation}}}
  \renewcommand{\)}{\end{equation}}}
\def\part#1{\frac{\partial}{\partial{#1}}}%
\def\group#1{\refstepcounter{equation}\setcounter{saveeqn}
 {\value{equation}}%
  \label{#1}\setcounter{equation}{0}%
\renewcommand{\theequation}{\mbox{\arabic{section}.\arabic{saveeqn}
\alph{equation}}}
  \renewcommand{\)}{\end{equation}}}
\newcommand{\reseteqn}{\setcounter{equation}{\value{saveeqn}}%
  \renewcommand{\theequation}{\arabic{section}.\arabic{equation}}%
  \renewcommand{\)}{\end{equation}}}

\newcommand{\aalpheqn}{\setcounter{saveeqn}{\value{equation}}%
  \stepcounter{saveeqn}\setcounter{equation}{0}%
  \renewcommand{\theequation}{\mbox{
        \Alph{subsection}.\arabic{saveeqn}\alph{equation}}}
   \renewcommand{\)}{\end{equation}}}
\newcommand{\areseteqn}{\setcounter{equation}{\value{saveeqn}}%
  \renewcommand{\theequation}{\Alph{subsection}.\arabic{equation}}%
  \renewcommand{\)}{\end{equation}}}

\renewcommand{\thefootnote}{\alph{footnote}}
\renewcommand{\(}{\begin{equation}}
\renewcommand{\)}{\end{equation}}
\newcommand{\ba}{\begin{eqnarray}}
\newcommand{\ea}{\end{eqnarray}}
\renewcommand{\a}{\alpha}
\renewcommand{\b}{\beta}

\renewcommand{\sl}{{\sqrt{\lambda}}}

\newcommand{\cbp}{\mathop{\vtop{\ialign{##\crcr
   $\hfil\displaystyle{}\hfil$\crcr\noalign{\kern-13pt\nointerlineskip}
   \BIG{)}\hskip0pt\crcr\noalign{\kern3pt}}}}}
\newcommand{\pa}{\mathop{\vtop{\ialign{##\crcr

$\hfil\displaystyle{\oplus}\hfil$\crcr\noalign{\kern+1pt\nointerlineskip
}
   \hspace{.08in}$^{\alpha=0}$\hskip6pt\crcr\noalign{\kern3pt}}}}}
\renewcommand{\hsp}{,\hspace{.3in}}
\newcommand{\p}{^\prime}
\newcommand{\pp}{^{\prime\prime}}

\catcode`\@=11
\def\vereq#1#2{\lower3pt\vbox{\baselineskip1.5pt \lineskip1.5pt
\ialign{$\m@th#1\hfill##\hfil$\crcr#2\crcr\sim\crcr}}}
\catcode`\@=12

\renewcommand{\(}{\begin{equation}}
\renewcommand{\)}{\end{equation}}

\def\vx{{\vec{x}}}

\def\vp{{\vec{p}}}

\def\vk{{\vec{k}}}

\def\kx#1{k_{#1x}}

\def\k#1{\k_{#1}}

\def\pin#1{\int \frac{d#1}{2\pi}}
\def\ppin#1{\int\hspace{-17pt}\sum \frac{d#1}{2\pi}^n}

\def\dint{\int\hspace{-12pt}\sum }

\def\pxinp{\int \frac{dp_x}{2\pi}}

\def\pxink{\int \frac{dk_x}{2\pi}}
\def\pyink{\int \frac{dk_y}{2\pi}}

\def\pinvp{\int \frac{d^{2}\vp}{(2\pi)^{2}}}
\def\pinvk{\int \frac{d^{2}\vk}{(2\pi)^{2}}}

\def\npinvk#1{\int \frac{d^{2}\vk_#1}{(2\pi)^{2}}}

\def\sinvk{\int\hspace{-17pt}\sum \frac{d^{2}\vk}{(2\pi)^{2}}}
\def\nsinvk#1{\int\hspace{-17pt}\sum \frac{d^{2}\vk_{#1}}{(2\pi)^{2}}}
\def\sxink{\int\hspace{-17pt}\sum \frac{dk_x}{2\pi}}
\def\nsxink#1{\int\hspace{-17pt}\sum \frac{dk_{x_{#1}}}{2\pi}}

\def\cc{\mathcal{C}}
\def\df{\mathcal{D}_{f}}

\def\avpd{a^\ddag_{\vec{p}}}

\def\avpm{a_{-\vec{p}}}

\def\omvk{\omega_{\vec{k}}}
\def\omvp{\omega_{\vec{p}}}

\def\nomvk#1{\omega_{\vec{k}_{#1}}}

\def\bdvk{B^{\ddagger}_{\vec{k}}}
\def\bvk{B_{\vec{k}}}

\def\bvkm{B_{-\vec{k}}}

\def\nbdvk#1{B^{\ddagger}_{\vec{k}_{#1}}} 
\def\nbvk#1{B_{\vec{k}_{#1}}}

\def\g{\mathfrak g}
\def\gvk{\g_{\vk}}

\def\ngvk#1{\g_{\vk_{#1}}}

\def\vkis{\vk_I^S}
\def\vkisx{k_{Ix}^S}
\def\vkisy{k_{Iy}^S}
\def\omvkis{\omega_{\vk_I^S}}
\def\vkias{\vk_I^{aS}}
\def\vkiasx{k_{Ix}^{aS}}
\def\vkiasy{k_{Iy}^{aS}}
\def\omvkias{\omega_{\vk_I^{aS}}}

\def\nomk#1{\omega_{k_{#1}}}

\def\omskys{\omega_{Sk_{Sy}}}

\def\red#1{\textcolor{red}{Jarah: #1}}
\def\blu#1{\textcolor{blue}{Hui: #1}}
\usepackage[dvipsnames]{xcolor}

\newcommand{\beas}{\begin{eqnarray*}}
\newcommand{\eeas}{\end{eqnarray*}}

\newcommand{\bquo}{\begin{quote}}
\newcommand{\enqu}{\end{quote}}

\def\lim#1{\stackrel{\rm{lim}}{{}_{#1}}}

\newcommand{\R}{{\mathbb R}}

\def\ok#1{\omega_{k_{#1}}}

\def\V#1{V^{(#1)}(\sqrt{\lambda}f(x))}

\def\ck{\csch\left(\frac{\pi k}{2\b}\right)}

\def\mb{\mathcal{B}}
\def\mc{\mathcal{C}}
\def\md{\mathcal{D}}
\def\me{\mathcal{E}}

\newcommand{\beq}{\begin{equation}}
\newcommand{\eeq}{\end{equation}}
\newcommand{\bea}{\begin{eqnarray}}
\newcommand{\eea}{\end{eqnarray}}

\newskip\humongous \humongous=0pt plus 1000pt minus 1000pt

\newif\ifdtup

\catcode`\@=11

\ifthenelse{\equal{\letter}{0}}{ 

\@addtoreset{equation}{section}
\def\theequation{\arabic{section}.\arabic{equation}}

\def\@normalsize{\@setsize\normalsize{15pt}\xiipt\@xiipt
\abovedisplayskip 14pt plus3pt minus3pt%
\belowdisplayskip \abovedisplayskip
\abovedisplayshortskip \z@ plus3pt%
\belowdisplayshortskip 7pt plus3.5pt minus0pt}

\def\small{\@setsize\small{13.6pt}\xipt\@xipt
\abovedisplayskip 13pt plus3pt minus3pt%
\belowdisplayskip \abovedisplayskip
\abovedisplayshortskip \z@ plus3pt%
\belowdisplayshortskip 7pt plus3.5pt minus0pt
\def\@listi{\parsep 4.5pt plus 2pt minus 1pt
      \itemsep \parsep
      \topsep 9pt plus 3pt minus 3pt}}

\relax

\def\section{\@startsection{section}{1}{\z@}{3.5ex plus 1ex minus  .2ex}{2.3ex plus .2ex}{\large\bf}}

\def\thesection{\arabic{section}}
\def\thesubsection{\arabic{section}.\arabic{subsection}}

\def\appendix{\setcounter{section}{0}
 \def\thesection{Appendix \Alph{section}}
 \def\thesubsection{\Alph{section}.\arabic{subsection}}
 \def\theequation{\Alph{section}.\arabic{equation}}}
\renewcommand{\theequation}{\arabic{section}.\arabic{equation}}

}{
\renewcommand{\theequation}{\arabic{equation}}

} 

\begin{document}
\def\thefootnote{\fnsymbol{footnote}}
\def\thetitle{(Anti-)Stokes Scattering on the Domain Wall String}
\def\autone{Hengyuan Guo}
\def\auttwo{Hui Liu}
\def\autthree{Jarah Evslin}

\def\affa{School of Physics and Astronomy, Sun Yat-sen University, Zhuhai 519082, China}
\def\affb{Yerevan Physics Institute, 2 Alikhanyan Brothers St., Yerevan 0036, Armenia}
\def\affc{Institute of Modern Physics, NanChangLu 509, Lanzhou 730000, China}
\def\affd{University of the Chinese Academy of Sciences, YuQuanLu 19A, Beijing 100049, China}



\ifthenelse{\equal{\pr}{1}}{
\title{\thetitle}
\author{\autone}
\author{\auttwo}
\author{\autthree}
\affiliation {\affa}
\affiliation {\affb}
\affiliation {\affc}

}{

\begin{center}
{\large {\bf \thetitle}}

\bigskip

\bigskip


{\large \noindent  
\autone{${}^{1}$} 
\footnote{guohy57@mail.sysu.edu.cn}, \auttwo{${}^{2}$}
\footnote{hui.liu@yerphi.am}
and \autthree{${}^{3,4}$} 
\footnote{jarah@impcas.ac.cn} 
}


\vskip.7cm

1) \affa\\
2) \affb\\
3) \affc\\
4) \affd\\

\end{center}

}

\begin{abstract}
\noindent
 In a general (2+1)-dimensional scalar model, we consider the scattering of a single quantum of radiation off a domain wall string, which excites or de-excites the wall's internal shape mode.  We refer to these two process as Stokes and anti-Stokes scattering.  We calculate the probability densities for these processes to first order in quantum field theory, as a function of incoming momenta and angles.  We include both the case in which the incident particle is transmitted and also that in which it bounces backward.  Our results are given as finite-dimensional integrals of normal modes and elementary functions, and numerical results are presented in the particular case of the $\phi^4$ double-well model.


\end{abstract}

%
\setcounter{footnote}{0}
\renewcommand{\thefootnote}{\arabic{footnote}}

\ifthenelse{\equal{\pr}{1}}
{
\maketitle
}{}

\section{Introduction}
\subsection{Motivation}
When an isolated string has a large radius of curvature as compared with its inverse width, lattice simulations show that it is well described by a Nambu-Goto effective action.  On the other hand, as soon as it interacts heavily with radiation or a network of strings, the ubiquitous small loops resulting from string reconnection in Nambu-Goto simulations are not present \cite{1703.06696}.  On the contrary, the emission of radiation, which is missing in the Nambu-Goto simulations, apparently prevents the formation of these small loops.  Furthermore, at least some of this radiation is at a frequency of order the inverse string width, much higher than the curvature of the string.

It was conjectured already in Ref.~\cite{0812.1929} that the radiation is the result of small scale structure in the strings, of order the string width.  For example, in the case of the $\phi^4$ double-well kink or domain wall soliton, there are two kinds internal modes available for such structure.  First, there is the zero-mode corresponding to translations.  Second, there is a shape mode corresponding to an oscillation of the string's width.  The latter is always at frequencies of order the inverse string width and indeed Ref.~\cite{2405.06030} found radiation at exactly twice the frequency of the internal shape mode, suggesting that pairs of shape mode quanta combine to escape to the bulk as in Ref.~\cite{mealberto}.

While this radiation has now been confirmed by numerous simulations, its origins are somewhat unclear as the microphysics of the interactions of a string with its environment is unclear.  Thus in Ref.~\cite{2006.13255} a program was initiated for understanding these interactions in lower-dimensional models, beginning with the kink in 1+1 dimensions and continuing to the 2+1 dimensional domain wall string in Refs.~\cite{2209.12945,2411.13521}.  These simulations of domain wall strings appear to show that the shape mode indeed interacts strongly with radiation in the bulk and so is likely to be important to the dynamics of string networks.  

To better understand this situation, in the present paper we present an analytic approach.  We consider interactions of a single quantum of radiation with a single quantum of a (2+1)-dimensional domain wall string's shape mode.  More precisely, we consider two processes, named after analogous processes in atomic physics \cite{stokes52,raman28,landsberg28}.  First, we consider Stokes scattering in which incoming radiation scatters off of a domain wall, perhaps reflecting, and in the process excites a shape mode.  Second, we consider anti-Stokes scattering in which incoming radiation collides with an excited string, de-exciting it.  

We compute the probability densities for these processes, as a function of the incoming and outgoing momenta, in the full quantum field theory with all radiative degrees of freedom.  However they are considered only at the leading order, where we expect that if applied to coherent states as in Ref.~\cite{mealberto} they will reproduce the corresponding results in classical field theory, which have however never been calculated to our knowledge.

\subsection{Background}
Consider a scalar field theory with a potential that has degenerate minima.  The $n$th based homotopy classes of the space of minima are the charges of the codimension $n+1$ topological solitons \cite{Kibble:1976sj}.  For example, in the case $n=0$, if there are two connected components of degenerate minima then there are two zeroth homotopy classes, one corresponding to a trivial soliton and the other to a codimension one soliton.  The theory will also contain perturbative excitations of the scalar field, which we will call mesons.

In 1+1-dimensions, the codimension one soliton is known as a kink, while in higher dimensions it is called a domain wall \cite{bog67,Zeldovich:1974uw,Dvali:1996xe,Steer:2006ik}. 
Motivated in part by cosmology \cite{muk24,dan24,big24} and in part by condensed matter applications \cite{yuki24,yxq}, there has recently been a resurgence in interest in domain wall phenomenology \cite{marra24,rout24,shiff24}.  General higher-dimensional stable solitons include (baby-)Skyrmions, monopoles and vortices which often lack exact analytical solutions~\cite{Manton:2004tk}. However some methods have been proposed to connect lower-dimensional integrable models to higher-dimensional models which give the exact domain wall solutions~\cite{Lou:2022ksp}.

In 1+1 dimensional classical field theory, one already observes a rich phenomenology of interactions between kinks and anti-kinks, revealing that different initial relative speeds yield a fractal structure characterized by resonance windows \cite{kklan,kk0,kk1,kk2,sfal21,car24}.  While initially it was believed that these windows result from the excitation of shape modes, the resonance windows have since been observed in kinks with no shape modes \cite{doreyf6,mech22,yunguo24} and so it has become clear that interactions with bulk excitations also play a critical role \cite{f622,nav23}. 

The spectral walls observed in classical field theory \cite{sw19,fermwall} become smoother in quantum field theory \cite{mechris}.  It has also been claimed that the lifetimes of oscillons are greatly reduced in quantum field theory \cite{hertzberg,vachosc} although this has been disputed in Ref.~\cite{noiosc}.  In general, it is fair to say that the lift of soliton phenomenology to quantum field theory remains poorly understood.

The quantization of 1+1-dimensional kinks has been extensively researched using various quantization methods such as those of Refs.~\cite{Dashen:1974ci,gjscc,Christ:1975wt}.  At one-loop there are simple tools, such as the classical-quantum correspondence of Ref.~\cite{vachcqc} and also spectral methods \cite{specrev} but more powerful tools tend to be cumbersome \cite{gj76} and even to lead to errors \cite{rebhan}.  Recently, linearized soliton perturbation theory (LSPT) has been introduced in Refs.~\cite{mekink,me2loop} which provides a lightweight and reliable tool.  

LSPT has been applied to kink-bulk interactions in Refs.~\cite{Evslin:2023ypw,Evslin:2022fzf}. In particular, at leading order a kink and a meson may interact via three inelastic scattering processes.  First there is meson multiplication, in which two mesons appear in the final state.  Next is Stokes scattering, in which a meson scatters off a ground state kink, exciting its shape mode.  Finally, anti-Stokes scattering is the process in which a meson scatters off an excited kink, leading to the de-excitation of its shape mode. LSPT has been used to calculate the probabilities of all three processes.  

Recently, linearized soliton perturbation theory has been generalized to 2+1-dimensional domain wall models, providing the domain wall string's ground state and reproducing the leading quantum correction to its tension \cite{medomain}, which was originally found in Ref.~\cite{Jaimungal}.

In the present paper, we will treat Stokes and anti-Stokes scattering on the domain wall string.  Unlike kink-meson scattering, in addition to the three processes above, domain wall-meson scattering admits two additional inelastic processes.  At leading order one may also excite or de-excite the zero-mode corresponding to the displacement of the domain wall.  This mode is not gapped, leading to the usual infrared divergences.  We will treat these separately in another work.  


Following a review of LSPT in Sec. \ref{revsez}, we calculate the probabilities of Stokes and anti-Stokes scattering as functions of the incoming meson momentum in Secs. \ref{stokessez} and \ref{asez} respectively. Our results are then specialized to the $\phi^4$ double-well model in Sec. \ref{fsez}.

\section{Linearized Soliton Perturbation Theory} \label{revsez}

\subsection{The Main Idea}

In Refs.~\cite{mekink,me2loop} a new Hamiltonian formalism, LSPT, has been introduced for calculations in the kink sector of a scalar quantum theory in 1+1 dimensions. In Ref.~\cite{medomain}, LSPT was generalized to the 2+1-dimensional case.

For simplicity, we consider a theory of a single scalar field $\phi(\vx)$ of mass $m$, whose perturbative excitations we call mesons.  We always work in the Scrodinger picture.  As reviewed in \ref{app}, we may decompose $\phi(\vx)$ and its dual momentum $\pi(\vx)$ as usual into creation and annihilation operators $a^\dag_\vp$ and $a_\vp$.  

\subsubsection{The Vacuum Sector}

We refer to the Fock space of mesons in the absence of a domain wall as the vacuum sector.  The vacuum sector states are built on the tree-level vacuum $|\Omega\rangle_0$,  which is defined as the state which is annihilated by all annihilation operators
\beq
a_\vp|\Omega\rangle_0=0. \label{omeq}
\eeq
More precisely, a basis of the vacuum sector is given by finite numbers of creation operators $a^\dag_\vp$ acting on the tree-level vacuum state $|\Omega\rangle_0$
\beq
|\vp_1\cdots \vp_n\rangle_0=\frac{a^\dag_{\vp_1}\cdots a^\dag_{\vp_n}}{\sqrt{(2\omega_{\vp_1})\cdots (2\omega_{\vp_n})}}|\Omega\rangle_0\hsp
\omega_\vp=\sqrt{m^2+|\vp|^2}. \label{vb}
\eeq
Hamiltonian eigenstates $|\Omega\rangle$ and $|\vp_1\cdots \vp_n\rangle$ in the vacuum sector can be constructed in perturbation theory, starting at leading order with $|\Omega\rangle_0$ and $|\vp_1\cdots \vp_n\rangle_0$ respectively.  

\subsubsection{The Domain Wall Sector}

The domain wall sector is the Fock space of states consisting of a quantum domain wall plus a finite number of fundamental perturbative mesons, including those bound to the wall.  The domain wall sector is built on the leading-order domain wall ground state, to which we now turn our attention.

Ordinary perturbation theory is a moment expansion in the field $\phi(\vx)$.  In the vacuum sector, these moments are small and so perturbation theory makes sense, at least as an asymptotic series at weak coupling.  On the other hand, if the domain wall is described by a classical soliton solution $\phi(\vx,t)=f(x)$
of the equations of motion, then the moments will be roughly those of $f(\vx)$, which contain powers of the inverse coupling.  Thus perturbation theory cannot be applied to domain wall sector states.

To solve this problem we introduce the displacement operator
\beq
\df={{\rm Exp}}\left[-i\int d^2\vx f(\vx)\pi(\vx)\right].
\eeq
This operator shifts the field by the classical solution
\beq
\phi(\vx)\df=\df\left(\phi(\vx)+f(\vx)\right).
\eeq
Although the moments involving the zero mode, which translates the wall, remain large, we can deal with these using translation invariance if the theory is translation-invariant, or else using wave packets.  As for other modes, we will now explain that this shift is just what we need to make their moments small and allow for a perturbative expansion.

Consider a state $|\psi\rangle$ in the vacuum sector.  Then $\df|\psi\rangle$ will be in the domain wall sector, up to a squeeze to which we will return to later.  In fact, all states in the domain wall sector can be constructed in this way.  Now the spectrum of Hamiltonian eigenstates in the domain wall sector consists of solutions to the equation
\beq
H\df|\psi\rangle=E\df|\psi\rangle. 
\eeq
This is a horrible, nonperturbative equation because $\df$ has inverse powers of the coupling, exponentiated.  Thus we could not hope to solve this equation in perturbation theory.  

However, acting with $\df^\dag$ on the left hand side one may rewrite this equation as
\beq
H\p|\psi\rangle=E|\psi\rangle\hsp H\p=\df^\dag H\df \label{df}
\eeq
where we call $H\p$ the soliton Hamiltonian.  Note that $H\p$ is unitarily equivalent to $H$ and so it has the same spectrum.  Furthermore, if $H$ is regularized then $H\p$ will be automatically regularized, and so there is no need for the regulator matching that was shown to lead to errors in Ref.~\cite{rebhan}.  Now the exponentiated inverse-coupling has disappeared.  

We will see below that this equation may be solved perturbatively.
In fact, in the presence of a domain wall, not only the Hamiltonian eigenstates, but also the form factors, and even amplitudes and probabilities for various processes can be calculated perturbatively.  For example, the time evolution of a domain wall sector state is given by
\beq
e^{-iHt}\df|\psi\rangle=\df e^{-iH\p t}|\psi\rangle
\eeq
and so while the state contains $\df$, one need only know how to act $H\p$ on the vacuum sector state $|\psi\rangle$.  The $\df$ prefactor is only applied at the end of the calculation.  

In fact, since $\df$ is unitary it will not affect matrix elements and so, without affecting any matrix elements, one may remove it by renaming all of our states.   More precisely, we will perform a passive transformation on all states, multiplying the kets used to name them by $\df^\dag$.  After this renaming of the states, we say that we are working in the domain wall frame.  In the domain wall frame, our tree-level vacuum is $\df^\dag|\Omega\rangle_0$ while the domain wall state above is just $|\psi\rangle$.  In the domain wall frame, the time-independent states are eigenstates of $H\p$ while the time evolution operator is $e^{-iH\p t}$.  We stress that, like all passive transformations, this renaming of the coordinates is just a convenience to remove the clutter of the $\df$ in every equation, and is not a physical operation on the states.


Because we had to choose a particular domain wall solution $f(x)$,  we have lost manifest translation invariance $f(x)\rightarrow f(x-x_0)$.   In perturbation theory we calculate locally, close to some base point in moduli space.  However at each step in perturbation theory we impose translation-invariance on the states,  so that our calculations in fact are valid everywhere.  This procedure is described in detail in Ref.~\cite{me2loop}.  This of course is in contrast with the collective coordinate method \cite{gjscc}, where $x_0$ is treated as a dynamical variable.  The collective coordinate method maintains manifest translation invariance, but pays a price \cite{gj76} as the commutators of $x_0$ are quite complicated.  This is the main difference between our methods.


\subsection{The Domain Wall Hamiltonian}
Now we turn to the solution of the eigenvalue equation (\ref{df}).  Continuing to work exclusively in the Schrödinger picture,   consider a (2+1)-dimensional scalar theory whose Hamiltonian is
\begin{equation}
H=\int d^2\vx: \mathcal{H}(\vx):_a \hsp
\mathcal{H}(\vx)=\frac{\pi^2(\vx)}{2}+\frac{(\partial_i \phi(\vx))^2}{2}
+\frac{V(\sqrt{\lambda} \phi(\vx))}{\lambda}
\end{equation}
where $V$ is the general potential with degenerate minima ($\phi^4$, the Sine-Gordon model, etc.).  In Sec.~\ref{fsez} we will specialize to a fixed, reflectionless potential.  This reflectionless potential is the only one that is renormalizable and bounded from below in 3+1 dimensions.  However, here in 2+1 dimensions, the $\phi^6$ theory is also renormalizable and so our results also apply to the domain walls of the $\phi^6$ theory.  Note that, the nonrenormalizability does not yet appear at one loop, and so much more general potentials may be treated as low energy effective theories.  

We name the coordinates $\vx=(x,y)$.  Consider the classical equations of motion for the classical field $\phi(\vx,t)$.  Choose $f(\vx)$ to be a fixed, stationary domain wall solution of these classical equations of motion which connects the two degenerate local minima.  Let the wall extend in the $y$ direction, so that $f$ only depends on $x$
\beq
\phi(\vx,t)=f(\vx)=f(x).
\eeq

The notation $::_a$ is the normal ordering of the creation and annihilation operators for plane waves.  It is defined at a mass scale $m$, for which we have two definitions
\beq
m^2=V^{(2)}(\sqrt{\lambda} f(\pm \infty))\hsp
V^{(n)}(\sqrt{\lambda} \phi(\vx))=\frac{\partial^n V(\sqrt{\lambda} \phi(\vx))}{(\partial \sqrt{\lambda} \phi(\vx))^n}
\eeq
corresponding to the scalar mass at the two minima of the potential at infinity.  If these disagree, then quantum corrections break the degeneracy and the domain wall becomes an accelerating false vacuum bubble wall \cite{wstabile}.  We will not consider this case.

To find the domain wall Hamiltonian, we will use the identity
\beq
:F(\phi,\pi):_a\df=\df :F(\phi+f,\pi):_a
\eeq
which holds with or without normal ordering.  We will find the eigenstates of $H\p$ perturbatively in the coupling constant $\lambda$.  To do this we decompose all quantities in powers of $\lambda$.  For example, the tension $\rho$ of the domain wall ground state is decomposed into $\sum_i \rho_i$ where each $\rho_i$ is of order $O(\lambda^{i-1})$.  Note that $\rho_0$ is just the classical tension of the classical domain wall string solution.

The domain wall Hamiltonian itself is decomposed into terms $H\p_i$ with $i$ factors of the fundamental fields when normal ordered.  Note that $H\p_i$ will be of order $O(\lambda^{-1+i/2})$.  The three terms $H\p_0$, $H\p_1$ and $H\p_2$ are not suppressed by any powers of $\lambda$, and so to arrive at the initial condition for our perturbation theory, all three need to be diagonalized exactly and simultaneously.  

Can we do this?  Fortunately the three operators are quite simple
\bea
H\p_0&=&Q_0\hsp H\p_1=0\\
H\p_2&=&\frac{1}{2}\int d\vx\left[:\pi^2(\vx):_a+:\left(\partial_x\phi(\vx)\right)^2:_a+:\left(\partial_y\phi(\vx)\right)^2:_a+V^{\prime\prime}[\g f(x)]:\phi^2(\vx):_a\right.].\nonumber
\eea
$H\p_0$ is a scalar and so is always diagonal, as is $H\p_1$.  We have therefore reduced the problem to diagonalizing $H\p_2$.  If we can do this, then fairly standard perturbative methods allow us to get the subleading corrections.

\subsection{Diagonalizing the Free Domain Wall Hamiltonian $H\p_2$}

The operator $H\p_2$ looks like the Hamiltonian of a free massive scalar, which is easily diagonalized by the basis (\ref{vb}), corresponding to operators that create plane waves.  However the mass term is position-dependent.  The basic problem is as follows.  In the case of the vacuum sector, the linearized equations of motion were solved by plane waves, and so the ground state at leading order in perturbation theory $|\Omega\rangle_0$ is the one with no plane waves (\ref{omeq}).  In the domain wall sector, on the other hand, the linearized equations of motion are solved by normal modes.  This suggests that we should somehow replace plane waves by normal modes.  We will now do this, following roughly the strategy in Ref.~\cite{cahill76}.

The linearized, classical equation of motion corresponding to $H\p_2$ is
\beq
\V{2}\phi(\vx,t)=(-\partial_t^2+\partial_x^2+\partial_y^2)\phi(\vx,t).
\eeq
Let us insert the constant-frequency Ansatz
\beq
\phi(\vx,t)=\gvk(\vx)e^{-i\omega_{\vk}t}=\g_{k_xk_y}(\vx)e^{-i\omega_{k_xk_y}t}
\eeq
where $\g_{k_xk_y}(\vx)$ is factorized as
\beq
\g_{k_xk_y}(\vx)=\g_{k_x}(x)e^{-ik_y y} .\label{facteq}
\eeq
Then $\g_{k_xk_y}(\vx)$ satisfies
\beq
\V{2}\g_{k_xk_y}(\vx)=\omega_{k_xk_y}^2\g_{k_xk_y}(\vx)+\g_{k_xk_y}\pp(\vx)  \hsp \omega_{k_xk_y}=\sqrt{m^2+k_x^2+k_y^2}\label{sl}
\eeq
or equivalently $\g_{k_x}(x)$ satisfies the Sturm-Liouville equation
\beq
\V{2}\g_{k_x}(x)=\omega_{k_x}^2\g_{k_x}(x)+\g^{''}_{k_x}(x) \hsp \omega_{k_x}=\sqrt{m^2+k_x^2}.
\eeq
Here $k_x$ is an abstract index which runs over all normal modes, unlike $k_y$, which is just a real number corresponding to the momentum in the $y$ direction.

There are three kinds of normal mode, classified by $\omega_{k_x}$.  First, there is a zero-mode
\beq
\g_B(x)=-\frac{f\p(x)}{\sqrt{\rho_0}}\hsp \omega_B=0
\eeq
corresponding to the broken translation symmetry.  Next, there is a continuum with $\omega_{k_x}\geq m$ corresponding to values of $k_x$ which are real numbers.  These continuum modes $\g_{k_x}(x)$ are asymptotically plane waves in the case of reflectionless domain walls, or more generally superpositions of right-moving and left-moving plane waves.  Finally, there may be shape modes $\g_S(x)$ within the mass gap, $ 0<\omega_{S}<m$.  The index $k_x$ is therefore rather abstract, it runs over the discrete value $B$, a discrete set of values labeled by the index $S$ and also all real numbers, which we continue to write as $k_x$.  The formulas above for $\omega$ in this sense may be interpreted as giving an imaginary value of $k_x$ in the case of bound modes, but as in Ref.~\cite{weigel24} this interpretation is not necessary, one may choose to use this formula only for the unbound modes while simply using the Sturm-Liouville equation as the definition of $\omega$ for the bound modes.

We take the bound modes to be real.  Fixing $k_x$ to be a continuum mode for a moment, our conventions are
\bea
&& \omega_{k_xk_y}=\sqrt{m^2+k_x^2+k_y^2}\hsp \omega_{Bk_y}=|k_y|\hsp \omega_{Sk_y}=\sqrt{\omega_{S}^2+k_y^2} \hsp \g^*_{\vk}(\vx)=\g_{-\vk}(\vx) \nonumber\\
&& \int dx \g_{B}(x)\g_{B}^{*}(x)=1\hsp \int dx \g_{S}(x)\g^*_{S\p}(x)=\delta_{SS\p}\nonumber\\
&&\int dx \g_{k_x}(x)\g^*_{k\p_x}(x)=2\pi\delta(k_x-k\p_x).\label{comp} 
\eea
In our abstract vector notation, the frequencies of the zero modes and shape modes may be written
\beq
\vk=(B,k_y)\hsp \omega_\vk=|k_y|
\eeq
and
\beq
\vk=(S,k_y)\hsp \omega_\vk=\sqrt{\omega_S^2+k_y^2}
\eeq
respectively.

The normal modes generate all bounded functions, as they solve a Sturm-Liouville equation.  Therefore, instead of plane waves, we may use them to decompose the fields \cite{medomain} 
\bea
\phi(\vx)&=&\pin{k_y}e^{-ik_y y}\left[\phi_{Bk_y}\g_B(x)+\sum_S \phi_{Sk_y}\g_S(x)+\pin{k_x}\phi_{k_xk_y}\g_{k_x}(x)
\right]\\
\pi(\vx)&=&\pin{k_y}e^{-ik_y y}\left[\pi_{Bk_y}\g_B(x)+\sum_S \pi_{Sk_y}\g_S(x)+\pin{k_x}\pi_{k_xk_y}\g_{k_x}(x)
\right].\nonumber
\eea
We then define an oscillator basis
\beq
\phi_{\vk}=B^\ddag_{\vk}+\frac{B_{-\vk}}{2\omega_{\vk}}\hsp
\pi_{\vk}=i\omega_{\vk}B^\ddag_{\vk}-i\frac{B_{-\vk}}{2}\hsp B_{\vk}^\ddag=\frac{B^\dag_\vk}{2\omega_\vk}
\eeq
where in the case of discrete modes we have defined
\beq
-(B,k_y)=(B,-k_y)\hsp -(S,k_y)=(S,-k_y).
\eeq
In other words, since the discrete normal modes are taken to be real and the negation of an index is actually complex conjugation, negation only affects the continuum modes.
This leads to the decomposition
\bea
\phi(\vx) &=&\phi_{0}\g_{B0}(\vx)+\sinvk\left(\bdvk+\frac{\bvkm}{2 \omvk}\right)\g_{\vk}(\vx) \nonumber\\
\pi(\vx) &=&\pi_{0} \g_{B0}(\vx)+i \sinvk\left(\omvk \bdvk-\frac{\bvkm}{2}\right)\gvk(\vx).\label{dec}
\eea
Here we have defined
\beq
\sinvk=\pyink\left(\pxink+\sum_{S}\right)+\int_{-\infty}^{-\epsilon} \frac{dk_y}{2\pi}+\int_{\epsilon}^{\infty} \frac{dk_y}{2\pi}
\eeq
where in the last two terms it is understood that $k_x=B$ is the zero mode.  These two terms are a kind of principal value prescription for potential infrared singularities as $k_y\rightarrow 0$ in the case of the translation mode $k_x=B$.  We have also defined the zero modes
\beq
\phi_0=\int_{-\epsilon}^{\epsilon} \frac{dk_y}{2\pi} \phi_{Bk_y}  e^{-ik_y y }\hsp \pi_0=\int_{-\epsilon}^{\epsilon} \frac{dk_y}{2\pi} \pi_{Bk_y}  e^{-ik_y y }.
\eeq

Here $\epsilon$ plays the role of an infrared cutoff for divergences that may occur at small $k_y$, as the translation mode is massless.  Such divergences may be expected as an arbitrarily small energy input may cause an arbitrarily large deformation of the string at large distances, a fact which is behind Coleman's theorem \cite{colemanth} stating that the translation mode is not a Goldstone boson on the string worldsheet.  In the present note we will consider only the leading order interactions between the unbound modes and the shape modes, which do not include the translation modes.  Therefore no infrared divergences will occur and we may safely set $\epsilon=0$, and so $\phi_0=\pi_0=0.$  However, at any finite coupling, there will be a long wavelength $1/k_y$ such that the probability of exciting one of these modes is of order unity, and so our perturbative truncation cannot be applied to the physics at arbitrarily large $|y|$.  In this note we will disregard this caveat, but will return to it in future work.




This provides a new basis of our operator algebra, and any operator may be written in terms of this new basis. The canonical commutation relations satisfied by $\phi(\vx)$ and $\pi(\vx)$ imply that they satisfy
\bea
\left[\phi_0, \pi_0\right]&=&2\epsilon i\hsp \left[B_{Bk_y},
B_{B k_y\p}^{\ddagger}\right]=2\pi\delta(k_y-k_y\p)\\
\left[B_{S k_y},B_{S\p k_y\p}^{\ddagger}\right]&=&\delta_{S S\p}2\pi\delta(k_y-k_y\p)
\hsp\left[B_{\vk_1}, B_{\vk_2}^{\ddagger}\right]=(2\pi)^2\delta^2\left(\vk_1-\vk_2\right).\nonumber
\eea
There is a correction to the $B_{Bk_y}$ commutator of order $\epsilon$ which we will neglect.

Finally we are ready to write $H_2\p$.  At $\epsilon=0$ it is \cite{medomain}
\beq
H\p_2=\int dy \rho_1 +\int\frac{dk_y}{2\pi}\left(|k_y|{B_{Bk_y}^{\ddag} B_{Bk_y}}+\sum_S\omega_{Sk_y} B_{Sk_y}^{\ddag} B_{Sk_y}\right)+\pinvk\omvk\bdvk\bvk.
\eeq

The one-loop contribution $\int dy \rho_1$ is infinite, however it is constant in the domain wall sector as we are expanding about a straight wall, and so we will ignore it.  

We recognize the remaining terms as a sum of oscillators for the normal modes, including the translation mode, the shape modes and the unbound mesons.  The leading order ground state of the string $\df\vac_0$ is therefore the simultaneous ground state of the operators
\beq
B_{Bk_y}\vac_0=B_{Sk_y}\vac_0=\bvk\vac_0=0. \label{v0}
\eeq

As the $B_\vk$ operators are linearly related to the $a_\vp$ operators, $\vac_0$ is related to $|\Omega\rangle_0$ by a Bogoliubov transform.  Therefore $\vac_0$ is a squeezed state and the string ground state $\df\vac_0$ is a squeezed coherent state.  As $\vac_0$ is an eigenstate of $H\p_2$, after perturbative corrections we may lift it to $\vac$ which will be an eigenstate of $H\p$ and so $\df\vac$ will be an eigenstate of $H$.  Indeed, it will be the lowest energy eigenstate in the wall sector.  On the contrary, $\df|\Omega\rangle$ is not an eigenstate of $H$ and the expectation value of its energy is higher than that of the ground state $\df\vac$.  At one-loop, the difference in energy densities is finite, unlike the case of 3+1 dimensions in which $\df|\Omega\rangle_0$ has an infinite energy density as compared with the vacuum while $\df\vac_0$ has a finite energy density \cite{noi31}. 

As described above, via a passive transformation we may simply drop the $\df$ from the left hand side of all states to remove clutter.  We will make this choice in the rest of the paper.  Therefore we will write the leading order domain wall state as the vector $\vac_0$ and time evolution will be generated by $H\p$.





\subsection{The Spectrum}

We have succeeded in diagonalizing $H\p_0+H\p_1+H\p_2$, the free part of the domain wall Hamiltonian.  The ground state is $\vac_0$.  

What about the excited states?  We may excite a shape mode with $y$-momentum by acting with $B^\ddag_{Sk_y}$
\beq
|Sk_y\rangle_0=B^\ddag_{Sk_y}\vac_0.
\eeq
Note that in our condensed notation, in which $\vk$ is an abstract vector whose first component may be a real number or a discrete index, we may alternatively write
\beq
\vk=(S,k_y)\hsp |\vk\rangle_0=B^\ddag_\vk\vac_0=|Sk_y\rangle_0.
\eeq
Physically the shape modes are internal excitations which change the profile of the wall, for example by making its width oscillate.  We may also excite translation modes
\beq
\vk=(B,k_y)\hsp |\vk\rangle_0=|Bk_y\rangle_0=B^\ddag_{Bk_y}\vac_0
\eeq
which are transverse waves, of wavenumber $k_y$ on the string, that oscillate the domain wall string back and forth in the $x$ direction.  There are also unbound mesons in the bulk
\beq
\vk=(k_x,k_y)\hsp|\vk\rangle_0=B^\ddag_{\vk}\vac_0.
\eeq
This is a state consisting of a domain wall plus a single unbound meson.  Our perturbative expansion, described in detail in Ref.~\cite{me2loop}, ensures that we rest in the center of mass frame.  However, in the domain wall's frame, the meson has momentum $\vk$.  

Finally, we may consider a domain wall with any number of bound and unbound excitations.  For example, we will be interested in states with both a bound shape mode and also an unbound mode
\beq
\vk_S=(S,k_y)\hsp |\vk_S\vk_1\rangle=B^\ddag_{\vk_1}B^\ddag_{Sk_y}\vac_0.
\eeq

These eigenstates of $H\p_0+H\p_1+H\p_2$ can be extended to eigenstates of the full $H\p$ via a perturbative expansion in $\lambda$ \cite{me2loop} using the higher order terms in the kink Hamiltonian
\beq
H\p_{n>2}=\lambda^{\frac{n}{2}-1}\int d^2\vx \frac{V^{(n)}(\sqrt{\lambda} f(x))}{n !}: \phi^n(\vx):_a.\label{hn}
\eeq
At each order, one first imposes translation invariance, which fixes the state up to a few coefficients, which can then be fixed using the eigenvalue equation for $H\p$.

\section{Stokes Scattering} \label{stokessez}

\subsection{The Initial Condition}

In a Stokes scattering event, one incoming meson of momentum $\vk_1$ is absorbed by a ground state domain wall.  The domain wall then emits one meson of momentum $\vk_2$ and the domain wall's shape mode is excited, traveling with a momentum $k_{Sy}$ along the wall. \par
The initial condition is therefore a superposition in momentum space
\beq
|\Phi\rangle_0=\npinvk1\a_{\vk_1}|\vk_1\rangle_0
\eeq
of one-meson states.  Recall that, as a result of our implicit $\df$, we are in the wall sector and so this state actually contains both a meson and also a ground state domain wall.

In coordinate space, the initial state is
\beq
|\Phi\rangle_0=\int d^2\vx\Phi(\vx)|\vx\rangle_0  \hsp |\vx\rangle_0=\pinvk\gvk(\vx)|\vk\rangle_0.
\eeq
Then the completeness relations and normalization conditions (\ref{comp}) on the normal modes imply
\bea
\Phi(\vx)&=&\pinvk\a_{\vk}\gvk^{\star}(\vx)  \hsp
\alpha_{\vk}=\int d^2\vx \Phi(\vx)\gvk(\vx)   \hsp
|\vk\rangle_0=\int d^2\vx\gvk^{\star}(\vx) |\vx\rangle_0.  \label{pach}
\eea

Choosing $f(x)$ such that the domain wall is centered at $x=0$, we will be interested in the following initial configuration.  The incoming meson begins in a wave packet of width $\sigma$ centered at the point $\vx=(x_0,y_0)$ with a momentum centered about $\vk_0=(k_{0x},k_{0y})$ satisfying
\beq
x_0 \ll -\frac{1}{m} \hsp 0<\frac{1}{k_{0x}},\frac{1}{m} \ll \sigma \ll |x_0|.
\eeq
The first condition implies that the meson begins far to the left of the wall.  The second implies that it is moving towards the wall with a wave packet that is broad enough that the meson is effectively monochromatic, yet narrow enough that the initial wave packet does not overlap with the wall.  The Gaussian wavepacket may be written
\beq
\Phi(\vx)=e^{-\frac{(x-x_0)^2}{4\sigma^2}+ix k_{0x}}e^{-\frac{(y-y_0)^2}{4\sigma^2}+ik_{0y}y}.
\eeq



The final states corresponding to Stoke's scattering consist of a meson and a domain wall whose shape mode is excited.  It is therefore a superposition of states of the form
\beq
\vk_S=(S,k_{Sy})\hsp |\vk_S\vk_2\rangle_0=B^\ddag_{Sk_{Sy}}\nbdvk2\vac_0. \label{fs}
\eeq
We use the subscript S to indicate that $k_{Sy}$ is the $y$ component of a shape mode's momentum $\vk_S$. 



\subsection{Time evolution at $O(\lambda^0)$}
 At leading order, the time evolution of the wave packet is
\bea
|\Phi(t)\rangle_{O(\lambda^0)}&=&e^{-iH\p_2t}|\Phi(0)\rangle_0=\pinvk\a_{\vk}e^{-i\omvk t}|\vk\rangle_0 \\
 &=&\int d^2\vx\pinvk\a_{\vk}e^{-i\omvk t}\gvk^{\star}(\vx) |\vx\rangle_0
 =\int d^2\vx\Phi(\vx,t)|\vx\rangle_0 \nonumber
\eea
where
\beq
\Phi(\vx,t)=\pinvk\a_{\vk}e^{-i\omvk t}\gvk^{\star}(\vx)
=\pinvk\int d^2\vx_1\Phi(\vx_1,t)\gvk(\vx_1)e^{-i\omvk t}\gvk^{\star}(\vx).
\eeq
Writing the resulting state in the position basis, one can see that it corresponds to a meson wave packet that moves to the right, passing through the wall without interaction.  The asymptotic forms of the normal modes can be used, if desired, to recover the usual phase shift associated with forward scattering.

\subsection{Time evolution at $O(\lambda^{1/2})$}
Using Eqs.~(\ref{hn}) and (\ref{a6}), we see that,  at order $O(\sqrt{\lambda})$, the only term in the domain wall Hamiltonian that can interpolate between our initial and final states is
\bea
H_I&=&\frac{\sqrt{\lambda}}{2}\int\frac{dk_{Sy}}{2\pi}\npinvk1\npinvk2\frac{V_{Sk_{Sy},\vk_2,-\vk_1}}{\nomvk1}B_{Sk_{Sy}}^{\ddagger}\nbdvk2\nbvk1\label{hint}\\
V_{Sk_{Sy},\vk_2,-\vk_1}&=&\int d^2\vx V^{(3)}(\sqrt{\lambda} f(\vx))\g_{Sk_{Sy}}(\vx)\g_{\vk_2}(\vx)\g_{-\vk_1}(\vx)=V_{S,k_{2x},-k_{1x}}(2\pi)\delta(k_{Sy}+k_{2y}-k_{1y})\nonumber\\
V_{S,k_2,-k_1}&=&\int dx V^{(3)}(\sqrt{\lambda}f(x))g_S(x)g_{k_2}(x)g_{-k_1}(x). \nonumber
\eea 
The corresponding terms in the time evolution operator are
\beq
e^{-it(H\p_2+H_I)}=e^{-itH\p_2}-i\int _0^t dt_1 e^{-i(t-t_1)H\p_2}H_I e^{-i t_1 H\p_2}+O(\lambda).
\eeq
We will drop the first term, corresponding to forward scattering, as it will not contribute to the matrix elements below.  Acting this on a state consisting of one domain wall and one meson, one finds a final state
\beq
\begin{aligned}
e^{-iH\p t}|\vk_1\rangle_0\bigg|_{O(\sl)}=&\frac{-i\sqrt{\lambda}}{2\nomvk1}\int\frac{dk_{Sy}}{2\pi}\npinvk2 e^{-\frac{it}{2}(\nomvk1+\omega_{\vk_S}+\nomvk2)}V_{Sk_{Sy},\vk_2,-\vk_1}\frac{\sin(\frac{\omega_{\vk_S}+\nomvk2-\nomvk1}{2})t}{(\omega_{\vk_S}+\nomvk2-\nomvk1)/2}|\vk_S\vk_2\rangle_0\\
=&\frac{-i\sqrt{\lambda}}{2\nomvk1}\int\frac{dk_{Sy}}{2\pi}\npinvk2 e^{-\frac{it}{2}(\nomvk1+\omega_{\vk_S}+\nomvk2)}V_{S,k_{2x},-k_{1x}}(2\pi)\delta(k_{Sy}+k_{2y}-k_{1y})\\
&\times\frac{\sin(\frac{\omega_{\vk_S}+\nomvk2-\nomvk1}{2})t}{(\omega_{\vk_S}+\nomvk2-\nomvk1)/2}|\vk_S\vk_2\rangle_0.
\end{aligned}
\eeq

Let us introduce the vector-valued function $\vk_I^S(\vk_2,\vk_S)$ with components
\beq
\vk_I^S=(k_{Ix}^S,k_{Iy}^S)
\eeq
where we leave the $\vk_2$ and $\vk_S$ dependence implicit.  The $S$ superscript in $k^S_I$ stands for ``Stokes" and not ``shape".  It is defined by
\beq
k_{Iy}^S=k_{2y}+k_{Sy}\hsp
k_{Ix}^S>0\hsp \omega_{\vk^S_I}=\omega_{\vk_2}+\omega_{\vk_S}.
\eeq
The last equality implies that this process is on-shell when $\vk_1=\vk_I^S$, while the first implies $k_{1y}=k_{Iy}^S$ as a result of $y$-momentum conservation among the mesons and shape modes. This $y$-momentum conservation can be seen in the interaction Hamiltonian (\ref{hint}).  Note that there is no $x$-momentum conservation among the mesons and shape modes, because the mesons and shape modes may transfer $x$-momentum to the domain wall.

 At large times, we may use the identity
\beq
\begin{aligned}
&\lim{t\rightarrow\infty}\frac{\sin[\left(\frac{\omega_{\vk_S}+\nomvk2-\nomvk1}{2}\right)t]}{(\omega_{\vk_S}+\nomvk2-\nomvk1)/2}(2\pi)\delta(k_{Sy}+k_{2y}-k_{1y})
=(2\pi)^2\delta(\omega_{\vk_S}+\nomvk2-\nomvk1)\delta(k_{Sy}+k_{2y}-k_{1y})\\
=&(2\pi)^2\delta(\omega_{\vec{k}_I^S}-\nomvk1)\delta(k_{Iy}^S-k_{1y})
=(2\pi)^2\left(\frac{{\omvkis}}{\vkisx}\right)\left(\delta(k_{1x}-\vkisx)+\delta(k_{1x}+\vkisx)\right)\delta(k_{1y}-\vkisy)\label{iden}
\end{aligned}
\eeq
where, in the last equality, we used the identity $\delta[g(x)]=\sum_i\frac{\delta(x-x_i)}{|g\p(x_i)|}$.  

Inserting this result into the wave packet (\ref{pach}), one finds the Stokes scattered part of the state at large times $t$
\beq
\begin{aligned}
&e^{-iH\p t}|\Phi\rangle_0\bigg|_{O(\sl)}\\
=&\npinvk1\frac{-i\sqrt{\lambda}\alpha_{\vk_1}}{2\nomvk1}\int\frac{dk_{Sy}}{2\pi}\npinvk2 V_{Sk_{Sy},\vk_2,-\vk_1}
e^{-\frac{it}{2}(\nomvk1+\omega_{\vk_S}+\nomvk2)}\frac{{\rm sin}\left[\left(\frac{\omega_{\vk_S}+\nomvk1-\nomvk2}{2}\right)t\right]}{(\omega_{\vk_S}+\nomvk2-\nomvk1)/2}|\vk_S\vk_2\rangle_0  \\
=&\frac{-i\sqrt{\lambda}}{2}\int\frac{dk_{Sy}}{2\pi}\npinvk2\frac{e^{-i{\omvkis} t}}{\vkisx}(\a_{\vkis}V_{S,k_{2x},-\vkisx}+\a_{-{\vkisx \vkisy}}V_{S,k_{2x},\vkisx})|\vk_S\vk_2\rangle_0. \label{sev}
\end{aligned}
\eeq
The meson wave packet begins far from the domain wall, where one may apply the asymptotic form of the normal modes
\bea
\gvk(\vx)&=&g_{k_x}(x)e^{-ik_yy}=\left\{\begin{tabular}{lll}
$(\mb_{k_x}e^{-ik_x x}+\mc_{k_x}e^{ik_x x})e^{-ik_yy}$  &\rm{if} & $x\ll  -1/m$\\
$(\md_{k_x}e^{-ik_x x}+\me_{k_x}e^{ik_x x})e^{-ik_yy} $  &\rm{if} & $x\gg 1/m$\\
\end{tabular}
\right. \label{gk}\\
|\mb_{k}|^2+|\mc_{\vk}|^2&=&|\md_{k}|^2+|\me_{k}|^2=1\hsp
\mb^*_{k}=\mb_{-k}\hsp
\mc^*_{k}=\mc_{-k}\hsp
\md^*_{k}=\md_{-k}\hsp
\me^*_{k}=\me_{-k}\nonumber
\eea
to evaluate the coefficients $\alpha_{\vk}$ of the wave packet.  As $\vkisx$ is defined to be positive and $k_{0x}$ is also chosen to be positive, in Eq.~(\ref{sev}) only two cases appear
\bea
\alpha_{{\vkis}}&=&(2\sigma\sqrt{\pi})^2\left[\mb_{\vkisx}e^{ix_0(k_{0x}-{\vkisx})-\sigma^2(k_{0x}-{\vkisx})^2}
+\mc_{\vkisx}e^{ix_0(k_{0x}+{\vkisx})-\sigma^2(\kx0+{\vkisx})^2}
\right]e^{-\sigma^2(k_{0y}-{\vkisy})^2+iy_0(k_{0y}-{\vkisy})}\nonumber\\
&=&(2\sigma\sqrt{\pi})^2\mb_{{\vkisx}}e^{ix_0(\kx0-\vkisx)-\sigma^2(\kx0-{\vkisx})^2} e^{-\sigma^2(k_{0y}-{\vkisy})^2+iy_0(k_{0y}-{\vkisy})}
\eea
and
\beq
\begin{aligned}
  \alpha_{-{\vkisx\vkisy}}=&(2\sigma\sqrt{\pi})^2\left[\mb^*_{\vkisx}e^{ix_0(\kx0-\vkisx)-\sigma^2(\kx0+\vkisx)^2}+\mc^*_{{\vkisx}}e^{ix_0(\kx0-{\vkisx})-\sigma^2(\kx0-\vkisx)^2}
\right]e^{-\sigma^2(k_{0y}-{\vkisy})^2+iy_0(k_{0y}-{\vkisy})}\\
=&(2\sigma\sqrt{\pi})^2\mc^*_{{\vkisx}}e^{ix_0(\kx0-{\vkisx})-\sigma^2(\kx0-{\vkisx})^2}e^{-\sigma^2(k_{0y}-{\vkisy})^2+iy_0(k_{0y}-{\vkisy})}.
\end{aligned}
\eeq
One finds the relevant part of the state at large times $t$
\bea
e^{-iH\p t}|\Phi\rangle_0\bigg|_{O(\sl)}
&=&-i\sigma\sqrt{\pi\lambda}\int\frac{dk_{Sy}}{2\pi}\npinvk2 e^{ix_0(k_{0x}-\vkisx)}e^{-\sigma^2(k_{0x}-\vkisx)^2}e^{-i{\omvkis}t}\nonumber\\
&&\times e^{-\sigma^2(k_{0y}-{\vkisy})^2+iy_0(k_{0y}-{\vkisy})}\left(\frac{\tilde{V}_{S,k_{2x},-\vkisx}}{\vkisx}\right)|\vk_S \vk_2\rangle_0\nonumber\\
\tilde{V}_{S,k_{2x},-\vkisx}&=&\mb_{\vkisx}V_{S,k_{2x},-\vkisx}+\mc^*_{\vkisx}V_{S,k_{2x},\vkisx}.  
\eea
Note that in the case of a reflectionless domain wall, $\mc=0$ and so $\left|\tilde{V}\right|=\left|V\right|$.

Let us consider another abstract vector
\beq
\vk_{S}\p=(S,k\p_{S y})
\eeq
corresponding to a shape mode with momentum $k\p_{Sy}$ along the kink.
Using the inner product
\beq
{}_0\langle \vk\p_S \vk_1|\vk_S\vk_2\rangle_0=\frac{(2\pi)^3\delta(\vk_1-\vk_2)\delta(k_{Sy}-k\p_{Sy})}{4\omega_{\vk_S}\nomvk2} {}_0\langle 0\vac_0
\eeq
we find the matrix elements
\beq
\frac{{}_0\langle \vk_S \vk_2|e^{-iH\p t}|\Phi\rangle_0}{{}_0\langle 0\vac_0}=\frac{-i\sigma^2\pi\sqrt{\lambda}e^{-\sigma^2(k_{0y}-{\vkisy})^2+iy_0(k_{0y}-{\vkisy})}}{2\omega_{\vk_S}\nomvk2 \vkisx} e^{ix_0(k_{0x}-\vkisx)}e^{-\sigma^2(k_{0x}-\vkisx)^2}e^{-i{\omvkis}t}\tilde{V}_{S,k_{2x},-\vkisx}
\eeq
which square to
\bea
\left|\frac{{}_0\langle \vk_S \vk_2|e^{-iH\p t}|\Phi\rangle_0}{{}_0\langle 0\vac_0}\right|^2
&=&\frac{\sigma^4\pi^2\lambda}{4\omega_{\vk_S}^2\nomvk2^2(\vkisx)^2}\left|\tilde{V}_{S,k_{2x},-\vkisx}\right|^2e^{-2\sigma^2(k_{0x}-\vkisx)^2}e^{-2\sigma^2(k_{0y}-\vkisy)^2}\nonumber\\
&=&\frac{\sigma^2\pi^{3}\lambda}{8\omega_{\vk_S}^2\nomvk2^2(\vkisx)^2}\left|\tilde{V}_{S,k_{2x},-\vkisx}\right|^2\delta(\vkisx-k_{0x})\delta(k_{0y}-\vkisy).
\eea
The last equality holds in the limit $\sigma\rightarrow\infty$.
To calculate the Stokes scattering probability, we will need the projector $\mathcal{P}$ onto final states with an excited domain wall and a single meson
\beq
\mathcal{P}=\int d^2\vk_2\int dk_{Sy}\mathcal{P}_{\rm{diff}}(\vk_2,k_{Sy})\hsp
\mathcal{P}_{\rm{diff}}(\vk_2,k_{Sy})=\frac{4\omega_{\vk_S}\nomvk2}{(2\pi)^3}\frac{|\vk_S\vk_2\rangle_0 {}_0\langle \vk_S\vk_2|}{{}_0\langle 0\vac_0} .
\eeq


Using the inner product
\beq
\frac{{}_0\langle \vk_1|\vk_2\rangle_0}{{}_0\langle 0\vac_0}=\frac{(2\pi)^2\delta(\vk_1-\vk_2)}{2\nomvk1}
\eeq
one obtains the normalization of the initial state
\bea
\frac{{}_0\langle\Phi|\Phi\rangle_0}{{}_0\langle 0\vac_0}&=&\npinvk1\npinvk2\a_{\vk_1}\a^*_{\vk_2}\frac{{}_0\langle \vk_2|\vk_1\rangle_0}{{}_0\langle 0\vac_0}=\pinvk\frac{|\alpha_{\vk}|^2}{2\omvk}=\frac{1}{2\nomvk0}\pinvk|\a_k|^2 \nonumber\\
&=&\frac{1}{2\nomvk0}\pinvk\int d^2\vx_1\int d^2\vx_2 \gvk(\vx_1)\gvk^*(\vx_2)\Phi(\vx_1)\Phi^*(\vx_2)
\nonumber\\
&=&\frac{1}{2\omega_{\vk_0}}\int d^2\vx |\Phi(\vx)|^2
=\frac{\sigma^2\pi}{\omega_{\vk_0}}
\eea
where we used $\omvk \sim \nomvk0$ in the last step on the first line.

Both ${}_0\langle \vk_1|\vk_2\rangle_0$ and ${}_0\langle 0\vac_0$ are infinite, and so the previous expression is strictly speaking not defined.  In Ref.~\cite{menorm}, we describe how such inner products may be calculated systematically by dividing the numerator and denominator by the translation group.  There are corrections with respect to the naive manipulations above, as a result of the nondiagonal action of the translation operator in the domain wall frame.  However, these corrections are always subleading by a power of $\sl$ and so do not affect our probability at $O(\lambda)$.

Finally we may assemble all of these ingredients to write the total probability $P_{\rm{S}}(\vk_0)$ and probability density $P_{\rm{S}}(\vk_0, k_{Sy})_{\rm{diff}}$ of Stokes scattering at $O(\lambda)$

\beq
\begin{aligned}
P_{\rm{S}}(\vk_0)
&=\frac{{}_0\langle\Phi|e^{iH\p t}\mathcal{P}e^{-iH\p t}|\Phi\rangle_0}{{}_0\langle\Phi|\Phi\rangle_0}
=\int\frac{dk_{Sy}}{2\pi}\npinvk2 \frac{4\omega_{\vk_S}\nomvk2}{{}_0\langle 0\vac_0}\frac{\left|{}_0\langle \vk_S\vk_2|e^{-iH\p t}|\Phi\rangle_0\right|^2}{{}_0\langle\Phi|\Phi\rangle_0}\\
&
=\int dk_{Sy} P_{\rm{S}}(\vk_0,k_{Sy})_{\rm{diff}}\\
P_{\rm{S}}(\vk_0,k_{Sy})_{\rm{diff}}&=\npinvk2 4\omega_{\vk_S}\nomvk2\frac{\sigma^2\pi^2\lambda}{16\omega_{\vk_S}^2\nomvk2^2(\vkisx)^2} \left|\tilde{V}_{S,k_{2x},-\vkisx}\right|^2 \frac{\delta(\vkisx-k_{0x})\delta(\vkisy-k_{0y})}{\sigma^2\pi/\nomvk0}\\
=&\npinvk2\frac{\pi\lambda\nomvk0 }{4\omega_{\vk_S}\nomvk2(k_{0x})^2}\left|\tilde{V}_{S,k_{2x},-\vkisx}\right|^2
\delta(k_{0x}-\vkisx)\delta(k_{0y}-k^S_y-k_{2y})\\
=&\frac{\lambda\nomvk0 }{8\omega_{\vk_S}k_{0x}^2}\int\frac{dk_{2x}}{2\pi}\frac{\left|\tilde{V}_{S,k_{2x},-k_{0x}}\right|^2}{\omega_{k_{2x}(k_{0y}-k^S_y)}}\delta(k_{0x}-\vkisx)\\
=&\frac{\lambda\left(\left|\tilde{V}_{S,\sqrt{(\nomvk0-\omega_{\vk_S})^2-m^2-(k_{0y}-k_{Sy})^2},-k_{0x}}\right|^2
+\left|\tilde{V}_{S,-\sqrt{(\nomvk0-\omega_{\vk_S})^2-m^2-(k_{0y}-k_{Sy})^2},-k_{0x}}\right|^2\right)}{16\pi\omega_{\vk_S}k_{0x}\sqrt{(\nomvk0-\omega_{\vk_S}^2-m^2-(k_{0y}-k_{Sy})^2}}
.\label{psm}
\end{aligned}
\eeq
The probability is the sum of two terms.  The first is the probability that the emitted meson travels in the same direction as the initial meson, corresponding to inelastic forward scattering, while the second term is the probability that it travels in the opposite direction, corresponding to backward scattering.  

The appearance of $\omega_{\vk_S}$ in the denominator is interesting.  If instead of the shape mode, we considered the process in which the translation mode is created, then this frequency would be $|\vk_{By}|$.  This enjoys an infrared divergence as $\vk_{By}\rightarrow 0$.  As a result, the probability density has a simple pole at zero momentum for the emission of soft translation modes and the total probability has a logarithmic divergence.  Of course, such divergences for the soft emission of massless modes are common in quantum field theory and can be handled by the imposition of our infrared cutoff.  In some cases they cancel with contributions involving zero modes in the initial state.  We will not consider this process further.

In the initial and final states (\ref{pach}) and (\ref{fs}), the meson travels at a constant velocity $k_{0x}/\omega_{\vk_0}$ in the $x$ direction  when far from the domain wall. It resembles the 1+1 dimensional kink case in Ref.~\cite{Evslin:2023ypw}.  
However, as we will discuss when we turn to the example, as a result of the $y$ momentum of the zero-mode, the domain wall string is not gapped.

\section{Anti-Stokes Scattering} \label{asez}

In anti-Stokes scattering, a meson wave packet is incident on an excited domain wall string.  The domain wall de-excites and then re-emits the meson.  We let $\vk_1$ be the initial meson momentum.  The domain wall's excited shape mode begins in a wave packet centered at the origin with a momentum of $k_{Sy}$ parallel to the wall, which we again describe using the abstract vector
\beq
\vk_S=(S,k_{Sy}).
\eeq

\subsection{Initial Condition}

The initial state is thus
\beq
\left|\Phi\right\rangle_0=
\npinvk1\a_{\vk_1}\int\frac{dk_{Sy}}{2\pi}e^{-\sigma_0^{2}k_{Sy}^2}\left|\vk_S \vk_1\right\rangle_0\hsp
|\vk_S \vk_1\rangle_0=B^\ddag_{Sk_{Sy}} \nbdvk1\vac_0. \label{inita}
\eeq

This process is somewhat different from Stokes scattering because the incoming meson may miss the localized shape mode, in which case there will be no reaction.  More concretely, when the meson reaches the wall, it will be separated from the shape mode in the $y$ direction by some impact parameter $b$.  If $b\gtrsim\sigma_0$ then we expect the amplitude for this interaction to be suppressed.  By integrating over different values of the impact parameter, we could determine the cross section of the shape mode for this process.

We will instead consider a simpler regime.  We will take $\sigma_0$ so to be so large that the meson necessarily passes through the shape mode, although the shape mode will be quite diffuse and so we expect a $1/\sigma_0$ suppression in the final scattering probability.  While we ignore the spreading of the meson wave packet, which is proportional to $|x_0|$, it will provide the dominant lower bound for $\sigma_0$.  However we also assume that $\sigma_0>>1/m$.
As $\sigma_0$ is so large, $y_0$ will be inconsequential and so we set $y_0=0$.  For simplicity, in this section we also impose that the incoming meson travels perpendicular to the string and so $k_{0y}=0$.  In all, we consider the initial condition
\bea 
\alpha_{\vk_1}&=&\int d^2\vx \Phi(\vx)\ngvk1(\vx)\nonumber   \\
\Phi(\vx)&=&e^{-\frac{(x-x_0)^2}{4\sigma^2}+ix k_{0x}}e^{-\frac{y^2}{4\sigma^2}
}\hsp x_0 \ll -\frac{1}{m} \hsp \frac{1}{k_{0x}},\frac{1}{m} \ll \sigma \ll |x_0|.  \label{aspach}
\eea

\subsection{Time Evolution}

At $O(\sl)$,  using 
(\ref{a7}), the only term which can transform such a state into a one-meson, one de-excited string state is
\beq
\begin{aligned}
   H_I=&\frac{\sqrt{\lambda}}{4}\int\frac{dk_{Sy}}{2\pi}\npinvk1\npinvk2\frac{V_{-\vk_S,\vk_2,-\vk_1}}{\omega_{\vk_S} \nomvk1} \nbdvk2 B_{\vk_S} B_{\vk_1}  \\
H_I |\vk_{S} \vk_1\rangle_0=&\frac{\sqrt{\lambda}}{4}\npinvk2\frac{V_{-\vk_S,\vk_2,-\vk_1}}{\nomvk1}\left|\vk_2\right\rangle_0  
\end{aligned}
\end{equation} 
where $-\vk_S=(S,-k_{Sy})$. At leading order, a finite time evolution then yields
\beq
\begin{aligned}
 &e^{-iH\p t}|\vk_S\vk_1\rangle_0\bigg|_{O(\sl)}=\frac{-i\sqrt{\lambda}}{4\omega_{\vk_S}\nomvk1} \npinvk2 V_{-\vk_S,\vk_2,-\vk_1}e^{-\frac{it}{2}(\nomvk1+\omega_{\vk_S}+\nomvk2)}
\frac{\sin(\frac{\nomvk1+\omega_{\vk_S}-\nomvk2}{\nomvk1}t)}{(\nomvk1+\omega_{\vk_S}-\nomvk2)/2}|\vk_2\rangle_0\label{eva}\\
=&\frac{-i\sqrt{\lambda}}{4\omega_{\vk_S}\nomvk1} \npinvk2 V_{S,k_{2x},-k_{1x}}e^{-\frac{it}{2}(\nomvk1+\omega_{\vk_S}+\nomvk2)}2\pi\delta(-k_{Sy}+k_{2y}-k_{1y})
\frac{\sin(\frac{\nomvk1+\omega_{\vk_S}-\nomvk2}{\nomvk1}t)}{(\nomvk1+\omega_{\vk_S}-\nomvk2)/2}|\vk_2\rangle_0.
\end{aligned}
\eeq

Let us introduce the vector-valued function $\vk_I^{aS}(\vk_2,\vk_S)$ with components
\beq
\vk_I^{aS}=(k_{Ix}^{aS},k_{Iy}^{aS})
\eeq
where we leave the $\vk_2$ and $\vk_S$ dependence implicit.  The $aS$ superscript in $k^{aS}_I$ stands for ``anti-Stokes".  It is defined by
\beq
k_{Iy}^{aS}=k_{2y}-k_{Sy}\hsp
k^{aS}_{Ix}>0\hsp \omega_{\vk^S_I}=\omega_{\vk_2}-\omega_{\vk_S}.
\eeq
This process is only on shell if $k_{1x}=\pm \vkiasx$.

At large times, only the on-shell $\vk_1$ values contribute as
\beq
\lim{t\rightarrow\infty}\frac{\sin(\frac{\nomvk1+\omega_{\vk_S}-\nomvk2}{2}t)}{(\nomvk1+\omega_{\vk_S}-\nomvk2)/2}2\pi\delta(-k_{Sy}+k_{2y}-k_{1y})
=\frac{\omvkias}{|\vkiasx|}(2\pi)^2\left(\delta(k_{1x}-\vkiasx)+\delta(k_{1x}+\vkiasx)\right)\delta(k_{1y}-\vkiasy).
\eeq
Substituting this limit into Eq.~(\ref{eva}) and folding the result into our initial wave packet (\ref{inita}), we find the anti-Stokes scattered part of the state at time $t$
\beq
\begin{aligned}
 &e^{-iH\p t}|\Phi\rangle_0\bigg|_{O(\sl)}\\
 =&\int\frac{dk_{Sy}}{2\pi}\frac{-i\sqrt{\lambda}e^{-\sigma_0^{2}k_{Sy}^2}}{4\omega_{\vk_S}}\npinvk1 
\frac{\a_{\vk_1}}{\nomvk1}\npinvk2 V_{-\vk_S,\vk_2,-\vk_1}e^{-\frac{it}{2}(\nomvk1+\omega_{\vk_S}+\nomvk2)}\frac{{\rm sin}\left[\left(\frac{\nomvk1+\omega_{\vk_S}-\nomvk2}{2}\right)t\right]}{(\nomvk1+\omega_{\vk_S}-\nomvk2)/2}|\vk_2\rangle_0\\
=&\int\frac{dk_{Sy}}{2\pi}e^{-\sigma_0^2k_{Sy}^2}\frac{-i\sqrt{\lambda}}{4\omega_{\vk_S}}\npinvk2 e^{-i\nomvk2 t}\left(\frac{1}{\vkiasx}\right)\left(\a_{\vkias}V_{S,k_{2x},-\vkiasx}+\a_{-\vkiasx\vkiasy}
V_{S,k_{2x},\vkiasx}\right)|\vk_2\rangle_0\\
=&\int\frac{dk_{Sy}}{2\pi}e^{-\sigma_0^{2}k_{Sy}^2}\frac{-i\sigma^2\pi\sqrt{\lambda}}{\omega_{\vk_S}} \npinvk2 e^{ix_0(k_{0x}-\vkiasx)-\sigma^2(k_{0x}-\vkiasx)^2}e^{-\sigma^2{\vkiasy}^2} e^{-i\omvkias t}
\left(\frac{\tilde{V}_{S,k_{2x},-\vkiasx}}{\vkiasx}\right)|\vk_2\rangle_0\\
\end{aligned}
\eeq
which is summarized by the matrix elements
\beq
\begin{aligned}
  \frac{{}_0\langle \vk_2|e^{-iH\p t}|\Phi\rangle_0}{{}_0\langle 0\vac_0}
=&\int\frac{dk_{Sy}}{2\pi}e^{-\sigma_0^{2}k_{Sy}^2}\frac{-i\sigma^2\pi\sqrt{\lambda}}{2\omega_{\vk_S}\nomvk2\vkiasx} e^{ix_0(k_{0x}-\vkiasx)-\sigma^2(k_{0x}-\vkiasx)^2}e^{-\sigma^2{\vkiasy}^2} \\
&\times e^{-i\omvkias t}\tilde{V}_{S,k_{2x},-\vkiasx} \\
=&\int dk_{Sy}\frac{\delta(k_{Sy})}{2\sigma_0\sqrt{\pi}}\frac{-i\sigma^2\pi\sqrt{\lambda}}{2\omega_{\vk_S}\nomvk2\vkiasx} e^{ix_0(k_{0x}-\vkiasx)-\sigma^2(k_{0x}-\vkiasx)^2}e^{-\sigma^2{\vkiasy}^2} \\
&\times e^{-i\omvkias t}\tilde{V}_{S,k_{2x},-\vkiasx}.
\end{aligned}
\eeq
In the last equality, we used the fact that $\sigma_0$ is much larger than the $y$-extent of the meson wave packet when it strikes the wall to replace the $e^{-\sigma_0^2k^2_{Sy}}$ using the nascent $\delta$ function identity $e^{-\sigma_0^2 k^2}=\frac{\sqrt{\pi}}{\sigma_0}\delta(k)$. This restricts our integration to a monochromatic initial shape mode wave.  Indeed, it then suffices to consider the monochromatic parts of all of the momenta. 

The squared matrix element then simplifies to
\beq
\begin{aligned}
    \left|\frac{{}_0\langle \vk_2|e^{-iH\p t}|\Phi\rangle_0}{{}_0\langle 0\vac_0}\right|^2
=&\int dk_{Sy}\frac{\delta(k_{Sy})}{(2\sigma_0\sqrt{\pi})^2}\frac{\sigma^4\pi^2\lambda}{4\omega_{\vk_S}^2\nomvk2^2(\vkiasx)^2}\left|\tilde{V}_{S,k_{2x},-\vkiasx}\right|^2e^{-2\sigma^2(k_{0x}-\vkiasx)^2-2\sigma^2{\vkiasy}^2} \\
=&\int dk_{Sy}\frac{\delta(k_{Sy})}{4\sigma_0^2\pi}\frac{\sigma^2\pi^3\lambda}{8\omega_{\vk_S}^2\nomvk2^2(\vkiasx)^2}
\left|\tilde{V}_{S,k_{2x},-\vkiasx}\right|^2\delta(\vkiasy)\delta(\vkiasx-k_{0x}).
\end{aligned}
\eeq

We want to calculate the probability that the final state has one ground state domain wall and one meson.  Such states are preserved by the projector
\beq
\mathcal{P}=\int d^2\vk_2 \mathcal{P}_{\rm{diff}}(\vk_2)\hsp  \mathcal{P}_{\rm{diff}}(\vk_2)=\frac{1}{(2\pi)^2} \frac{2\nomvk2|\vk_2\rangle_0{}_0\langle \vk_2|}{{}_0\langle 0\vac_0} .
\eeq

Using the inner product
\beq
\frac{{}_0\langle \vk_1\vk_S|\vk_2\vk_S\p\rangle_0}{{}_0\langle 0\vac_0}=\frac{(2\pi)^3\delta(\vk_1-\vk_2)\delta(k_{Sy}-k_{Sy}\p)}{4\nomvk1\omega_{\vk_S}}
\eeq
we obtain the correction factor resulting from the norm of the initial state
\beq
\begin{aligned}
 \frac{{}_0\langle\Phi|\Phi\rangle_0}{{}_0\langle 0\vac_0}
=&\int\frac{dk_{Sy}\p}{2\pi}e^{-\sigma_0^{2}k_{Sy}^{'2}}\int\frac{dk_{Sy}}{2\pi}e^{-\sigma_0^{2}k_{Sy}^2}\npinvk1\npinvk2\alpha_{\vk_1}\alpha^*_{\vk_2}\frac{{}_0\langle Sk_{Sy}\vk_2|Sk_y^{S'}\vk_1\rangle_0}{{}_0\langle 0\vac_0}\\
=&\int\frac{dk_{Sy}}{2\pi}e^{-2\sigma_0^{2}k_{Sy}^2}\pinvk\frac{|\alpha_{\vk}|^2}{4\omega_{\vk_S}\omvk}
=\int\frac{dk_{Sy}}{2\pi}e^{-2\sigma_0^{2}k_{Sy}^2}\pinvk\frac{|\alpha_{\vk}|^2}{4\omega_{\vk_S}\omega_{\vk_0}}\\
=&\int\frac{dk_{Sy}}{2\pi}e^{-2\sigma_0^{2}k_{Sy}^2}\pinvk\int d^2\vx_1\int d^2\vx_2\frac{\gvk(\vx_1)\g^*_{\vk}(\vx_2)\Phi(\vx_1)\Phi^*(\vx_2)}{4\omega_{\vk_S}\omega_{\vk_0}} \\
=&\int\frac{dk_{Sy}}{2\pi}\frac{e^{-2\sigma_0^{2}k_{Sy}^2}}{4\omega_{\vk_S}\omega_{\vk_0}}\int d^2\vx |\Phi(\vx)|^2
=\frac{\sqrt{\pi}}{\sqrt{2}\sigma_0 }\frac{\sigma^2\pi}{2\omega_{S}\nomvk0}.
\end{aligned}
\eeq

Finally we may assemble all of these ingredients to write the total probability $P_{\rm{aS}}(\vk_0)$
of Stokes scattering at $O(\lambda)$ 
\beq
\begin{aligned}
&P_{\rm{aS}}(\vk_0)=\npinvk2 P_{\rm{aS}}(\vk_2,k_{Sy})_{\rm{diff}}
=\frac{{}_0\langle\Phi|e^{iH\p t}\mathcal{P}e^{-iH\p t}|\Phi\rangle_0}{{}_0\langle\Phi|\Phi\rangle_0}
=\npinvk2 \frac{2\nomvk2}{{}_0\langle 0\vac_0}\frac{\left|{}_0\langle\vk_2|e^{-iH\p t}|\Phi\rangle_0\right|^2}{{}_0\langle\Phi|\Phi\rangle_0}\\
=&\npinvk2 \int dk_{Sy}\frac{\delta(k_{Sy})}{4\sigma_0^2\pi}2\nomvk2\frac{\sigma^2\pi^3\lambda}{8\omega_{\vk_S}^2\nomvk2^2(\vkisx)^2} \left|\tilde{V}_{S,k_{2x},-\vkisx}\right|^2 \frac{\delta(\vkisx-k_{0x})\delta(\vkisy)}{\frac{\sqrt{\pi}}{\sqrt{2}\sigma_0}\frac{\sigma^2\pi}{2\omega_{S}\nomvk0}}\\
=&\int dk_{Sy}\frac{\delta(k_{Sy})}{\sqrt{2\pi}\sigma_0}
\npinvk2\frac{\pi^2\lambda\nomvk0 }{2\omega_{\vk_S}\nomvk2(k_{0x})^2}\left|\tilde{V}_{S,k_{2x},-\vkisx}\right|^2
\delta(k_{0x}-\vkisx)\delta(k_{Sy}-k_{2y})\\
=&\int dk_{Sy}\frac{\delta(k_{Sy})}{\sqrt{2\pi}\sigma_0}\frac{\pi\lambda\nomvk0 }{4\omega_{\vk_S}(k_{0x})^2}\int\frac{dk_{2x}}{2\pi}\frac{\left|\tilde{V}_{S,k_{2x},-k_{0x}}\right|^2}{\omega_{\vk_2=(k_{2x},k_{Sy}
)}}\delta(k_{0x}-\vkisx)\\
=&\frac{\lambda\left(\left|\tilde{V}_{S,\sqrt{(\nomvk0+\omega_{S})^2-m^2},-k_{0x}}\right|^2
+\left|\tilde{V}_{S,-\sqrt{(\nomvk0+\omega_{S})^2-m^2},-k_{0x}}\right|^2\right)}{8\sqrt{2\pi}\sigma_0 \omega_{S}k_{0x}\sqrt{(\nomvk0+\omega_{S})^2-m^2}}
.\label{pasm}
\end{aligned}
\eeq
 In the numerator of the last expression in Eq.~(\ref{pasm}), we note  again that the first term is the probability that the outgoing meson travels in the same direction as the incoming meson while and the second is the probability that the outgoing meson travels in the opposite direction as the incoming meson.


\section{Example: $\phi^4$ Double well Model} \label{fsez}

\subsection{Analytic Results}

Consider the $\phi^4$ double well model, which is defined by the potential
\beq
V(\sqrt{\lambda}\phi(\vx))=\frac{\lambda\phi^2(\vx)}{4}\left(\sqrt{\lambda}\phi(\vx)-\sqrt{2}m\right)^2
\eeq
and has a classical domain wall solution
\beq
\phi(\vx,t)=f(x)=\beta\sqrt\frac{2}{\lambda}\left[1+\tanh\left(\beta x 
\right)
\right]\hsp \beta=\frac{m}{2}.
\eeq
This domain wall string has a single shape mode, with frequency
\beq
\omega_{\vk_S}=\sqrt{\omega_S^2+k_{Sy}^2}\hsp \omega_S=\frac{\sqrt{3}m}{2}\hsp
\vk_S=(S,k_{Sy}).
\eeq
The normal modes as usual can all be factorized
\beq
\gvk(\vx)=g_{k_x}(x)e^{-ik_yy}
\eeq
where the $x$-dependent parts are
\bea
g_k(x)&=&\frac{e^{-ikx}}{\ok{} \sqrt{k^2+\b^2}}\left[k^2-2\b^2+3\b^2\sech^2(\b x)-3i\b k\tanh(\b x)\right]\label{nmode}\\
g_S(x)&=&\sqrt\frac{3\b}{2}\tanh(\b x)\sech(\b x)\hsp
g_B(x)=\frac{\sqrt{3\b}}{2}\sech^2(\b x).\nonumber
\eea
From these we find the three-point coupling
\beq
V_{Sk_1k_2}=\pi\frac{3\sqrt{3}}{8}\frac{\left(17\b^4-(\ok{1}^2-\ok{2}^2)^2\right)(\b^2+k_1^2+k_2^2)+8\b^2k_1^2k_2^2}{\b^{3/2}\ok{1}\ok{2}\sqrt{\b^2+k_1^2}\sqrt{\b^2+k_2^2}}\sech\left(\frac{\pi(k_1+k_2)}{2\b}\right).\label{v12}
\eeq


For simplicity, we will consider only the case $k_{0y}=0$ of an incoming meson whose wave packet moves perpendicular to the domain wall string
\beq
\vk_0=(k_{0x},0)
\eeq
and we will consider the shorthand notation $k_0=k_{0x}$.
Now the three-point interaction $\tilde{V}$ that appears in (\ref{psm}) becomes
\beq
\begin{aligned}
&\tilde{V}_{S,\pm\sqrt{(\omega_{k_0}-\omega_{\vk_S})^2-m^2-k_{Sy}^2},
-k_0}
=V_{S,\pm\sqrt{(\omega_{k_0}-\omega_{\vk_S})^2-m^2-k_{Sy}^2},
-k_0}\\
=&\pi\frac{3\sqrt{3}}{8}\frac{(8\b^4+12\b^2\omega_{k_0}\omega_{\vk_S}-4\nomk0^2\omega_{\vk_S}^2)(\nomk0^2-2\nomk0\omega_{\vk_S}+k_0^2)+8\b^2(\nomk0^2-2\nomk0\omega_{\vk_S}-\b^2)k_0^2}
{\b^{3/2}\sqrt{(\nomk0-\omega_{\vk_S})^2-k_{Sy}^2}\nomk0\sqrt{(\nomk0-2\omega_{\vk_S})\nomk0}\sqrt{\b^2+k_0^2}}\\
&\times\sech\left(\frac{\pi(\pm\sqrt{(\nomk0-\omskys)^2-m^2-k_{Sy}^2}-k_0)}{2\b}\right)\\
=&\frac{3\pi\sqrt{3\nomk0}}{\b^{3/2}}
\frac{\b^2(\b^2+k_0^2+\nomk0\omega_{\vk_S})-\omega_{\vk_S}^2(\b^2+\nomk0^2-\nomk0\omega_{\vk_S})}{\sqrt{\nomk0^2+3\b^2-2\nomk0\omega_{\vk_S}}\sqrt{(\nomk0-2\omega_{\vk_S})}\sqrt{\nomk0^2-3\b^2}}\\
&\times\sech\left(\frac{\pi(\pm\sqrt{\nomk0^2-\b^2-2\nomk0\omega_{\vk_S}}-k_0)}{2\b}\right).\label{vs}
\end{aligned}
\eeq
The first equality results from the fact that this potential is reflectionless.

In the case of Stokes scattering, the conservation of energy implies that $\omega_{k_0}>\omega_{\vk_S}$.  As it is a square, $k_{2x}^2=(\omega_{k_0}-\omega_{\vk_S})^2-m^2-k_{Sy}^2$ is also nonnegative.  Thus, for a fixed $k_0$, we obtain two upper bounds on the resulting shape mode momentum $|k_{Sy}|$ 
\beq
|k_{Sy}| \leq\sqrt{k_0^2+\frac{m^4}{4}}\hsp
|k_{Sy}|\leq\frac{\sqrt{16k_0^4-24k_0^2m^2-39m^4}}{8\sqrt{m^2+k_0^2}}.\label{pskys}
\eeq
It is easy to see that the second bound is stronger than the first bound.

In particular, this second limit implies that Stokes scattering is only possible if
\beq
16k_0^4-24k_0^2m^2-39m^4 \geq 0 
\rightarrow k_0\geq \sqrt{\frac{1}{4}(3m^2+4\sqrt{3}m^2)}.\label{psk0c}
\eeq

For a fixed $k_{Sy}$ in the allowed interval, the probability density of Stokes scattering is then
\beq
\begin{aligned}
&P_{\rm{S}}(k_0,k_{Sy})_{\rm{diff}}=\frac{\lambda\left(\left|\tilde{V}_{S,\sqrt{(\nomk0-\omega_{\vk_S})^2-m^2-k_{Sy}^2},
-k_{0}}\right|^2+\left|\tilde{V}_{S,-\sqrt{(\nomk0-\omega_{\vk_S}})^2-m^2-k_{Sy}^2,-k_{0}}\right|^2\right)}
{16\pi k_0\omega_{\vk_S}\sqrt{(\nomk0-\omega_{\vk_S})^2-m^2-k_{Sy}^2}}\\
=&\frac{27\lambda\pi\nomk0}{16\omega_{\vk_S}k_0\sqrt{(\nomk0-\omega_{\vk_S})^2-4\b^2-k_{Sy}^2}}
\frac{\left(\b^2(\b^2+k_0^2+\nomk0\omega_{\vk_S})-\omega_{\vk_S}^2(\b^2+\nomk0^2-\nomk0\omega_{\vk_S})\right)^2}
{\b^3(\nomk0^2+3\b^2-2\nomk0\omega_{\vk_S})(\nomk0-2\omega_{\vk_S})(\nomk0^2-3\b^2)}\\
&\times\left[\sech^2\left(\frac{\pi(\sqrt{\nomk0^2-\b^2-2\nomk0\omega_{\vk_S}}-k_0)}{2\b}\right)
+\sech^2\left(\frac{\pi(-\sqrt{\nomk0^2-\b^2-2\nomk0\omega_{\vk_S}}-k_0)}{2\b}\right)\right].
\end{aligned}
\eeq
As always, the first sech term represents the probability that the outgoing meson continues to the right, corresponding to inelastic forward scattering.  The second sech term indicates the probability that it reflects.   We note however that the outgoing meson also acquires momentum in the $y$ direction. 
The ratio of these two probabilities corresponds directly to the ratio of the two sech terms.
The total probability $ P_{\rm{S}}(k_0)$ of Stokes scattering is 
\beq
\begin{aligned}
P_{\rm{S}}(k_0)=&\int_{-k_{y max}^S}^{k_{y max}^S} dk_{Sy} P_{\rm{S}}(k_0,k_{Sy})_{\rm{diff}} 
\hsp k_0 \geq \sqrt{\frac{1}{4}(3m^2+4\sqrt{3}m^2)}\\
k_{y max}^S=&\frac{\sqrt{16k_0^4-24k_0^2m^2-39m^4}}{8\sqrt{m^2+k_0^2}}.
\end{aligned}
\eeq

Although we have not provided the total probability in the case $k_{0y}\neq 0$, we have calculated it and found that in fact the probability of Stokes scattering is independent of $k_{0y}$.  This is a result of the invariance of our domain wall ground state under boosts in the $y$ direction.


Similarly, in the case of anti-Stokes scattering
\beq
\begin{aligned}
&\tilde{V}_{S,\pm\sqrt{(\omega_{k_0}+\omega_{S})^2-m^2},-k_0}
=V_{S,\pm\sqrt{(\omega_{k_0}+\omega_{S})^2-m^2},-k_0}  \\
=&\frac{3\sqrt{3}\pi}{8}\frac{(8\b^4-12\b^2\nomk0\omega_{S}-4\nomk0^2\omega_{S}^2)(\nomk0^2+2\nomk0\omega_{S}+k_0^2)+8\b^2 k_0^2(\nomk0^2+2\nomk0\omega_{S}-\b^2)}{\b^{3/2}\sqrt{(\omega_{k_0}+\omega_{S})^2}\nomk0\sqrt{(\omega_{k_0}+2\omega_{S})\nomk0}\sqrt{\b^2+k_0^2}}\\
&\times \sech\left(\frac{\pi(\pm\sqrt{(\nomk0+\omega_S)^2-m^2}-k_0)}{2\b}\right)\\
=&\frac{3\pi\sqrt{3\nomk0}}{\b^{3/2}}
\frac{\b^2(\b^2+k_0^2-\nomk0\omega_{S})-\omega_{S}^2(\b^2+\nomk0^2+\nomk0\omega_{S})}{\sqrt{\nomk0^2+3\b^2+2\nomk0\omega_{S}}\sqrt{\nomk0+2\omega_{S}}\sqrt{\nomk0^2-3\b^2}}\times\sech\left(\frac{\pi(\pm\sqrt{\nomk0^2-\b^2+2\nomk0\omega_{S}}-k_0)}{2\b}\right)\\
\end{aligned}
\eeq
leading to the probability 
\beq
\begin{aligned}
&P_{\rm{aS}}(k_0)=\frac{\lambda\left(\left|\tilde{V}_{S,\sqrt{(\nomk0+\omega_S)^2-m^2},-k_{0}}\right|^2
+\left|\tilde{V}_{S,-\sqrt{(\nomk0+\omega_{S})^2-m^2},-k_{0}}\right|^2\right) }
{8\sqrt{2\pi}\sigma_0\omega_S k_0\sqrt{(\nomk0+\omega_{S})^2-4\b^2}}\\
=&\frac{27\lambda\pi^{\frac{3}{2}}\nomk0}{8\sqrt{2}\sigma_0\omega_{S}k_0\sqrt{(\nomk0+\omega_{S})^2-4\beta^2}}
\frac{\left(\b^2(\b^2+k_0^2-\nomk0\omega_{S})-\omega_{S}^2(\b^2+\nomk0^2+\nomk0\omega_{S})\right)^2}
{\b^3(\nomk0^2+3\b^2+2\nomk0\omega_{S})(\nomk0+2\omega_{S})(\nomk0^2-3\b^2)}\\
&\times\left[\sech^2\left(\frac{\pi(\sqrt{\nomk0^2-\b^2+2\nomk0\omega_{S}}-k_0)}{2\b}\right)
+\sech^2\left(\frac{\pi(-\sqrt{\nomk0^2-\b^2+2\nomk0\omega_{S}}-k_0)}{2\b}\right)\right].
\end{aligned}
\eeq

\subsection{Numerical Results }

The probability densities depend on the dimensionless coupling
$\frac{\lambda}{m}$ as well as the dimensionless momenta $k_0/m$ and $k_{Sy}/m$.  We have fixed our units so that the meson mass, far from a domain wall, is $m=1$.  Then the threshold incoming momentum (\ref{psk0c}) for Stokes scattering becomes
\beq
k_0\geq \sqrt{\frac{1}{4}(3m^2+4\sqrt{3}m^2)} =1.57545. \label{k0m}
\eeq
For each value $k_0$ of the incoming momentum, the resulting shape mode momentum $ k_{Sy}$ is  restricted to region $|k_{Sy}|\leq \sqrt{k_0^2+\frac{5}{4}-2\sqrt{1+k_0^2}}$. 

\begin{figure}[htbp]
\centering
\includegraphics[width = 0.45\textwidth]{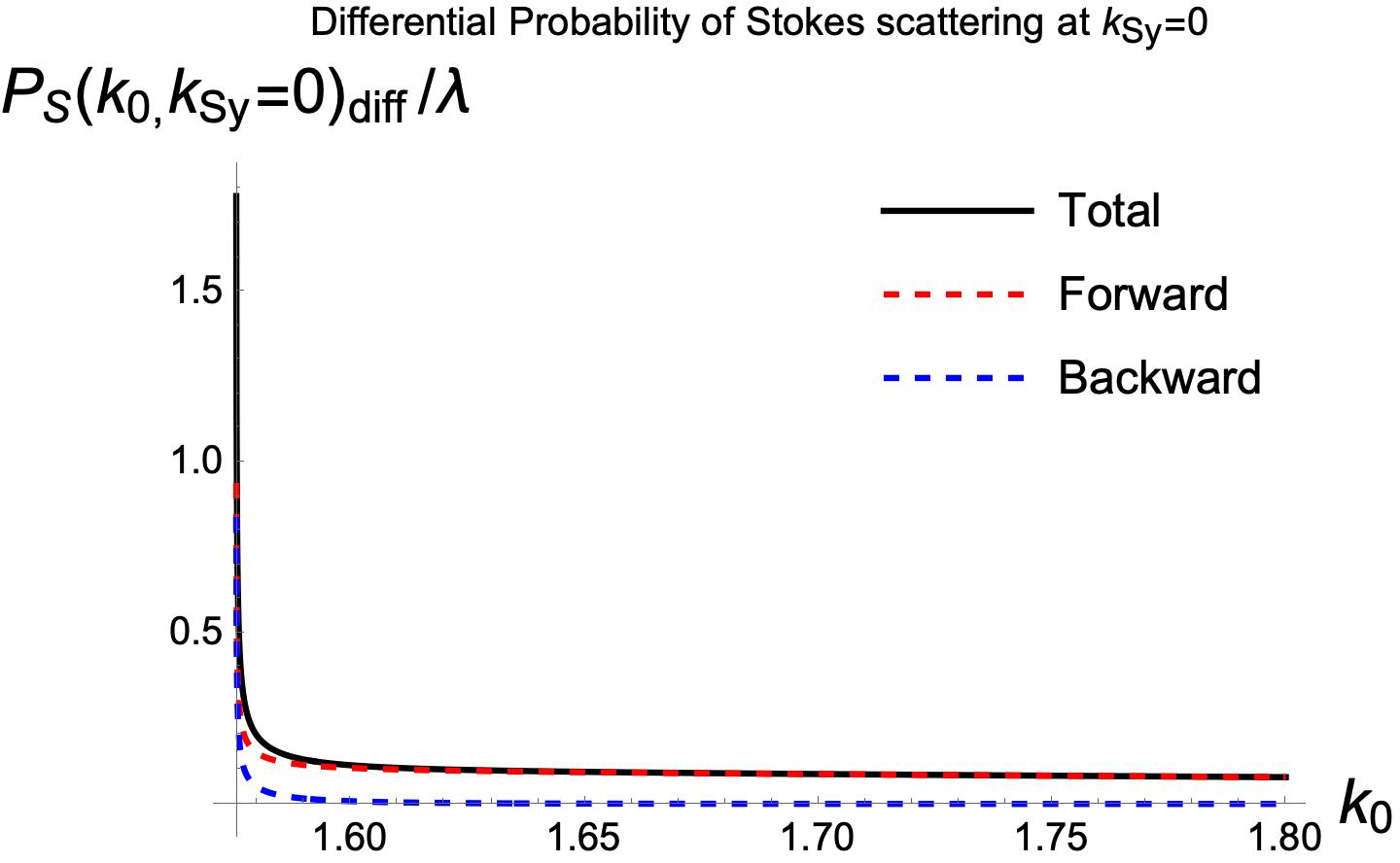}
\includegraphics[width = 0.45\textwidth]{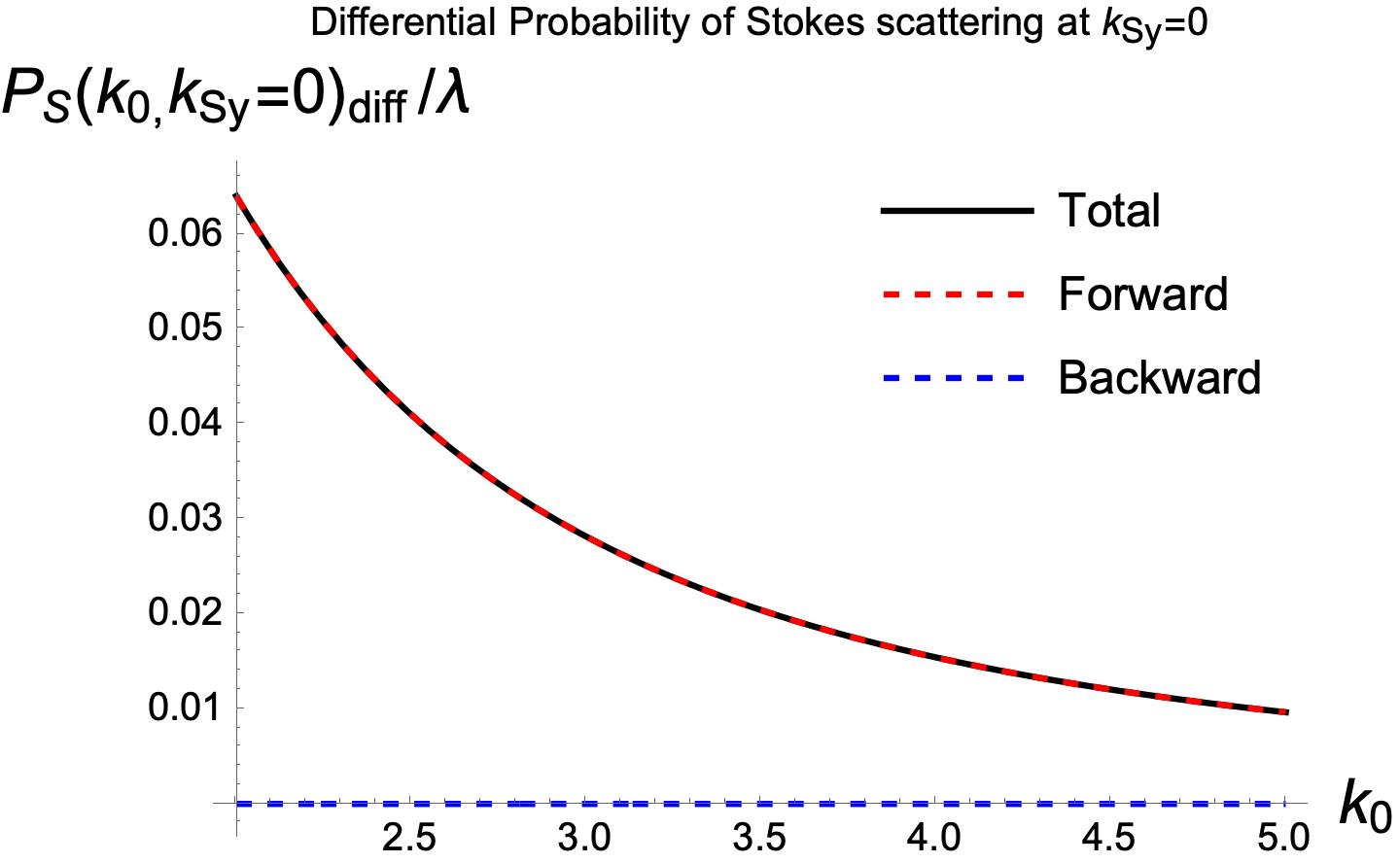}
\caption{The forward , backward  and total probabilities $P_{S}(k_0)_{\rm{diff}}$ of Stokes scattering, with $m=1$.  The region near the threshold is shown in the left panel.} \label{2dps1}
\end{figure}

\begin{figure}[htbp]
\centering
\includegraphics[width = 0.45\textwidth]{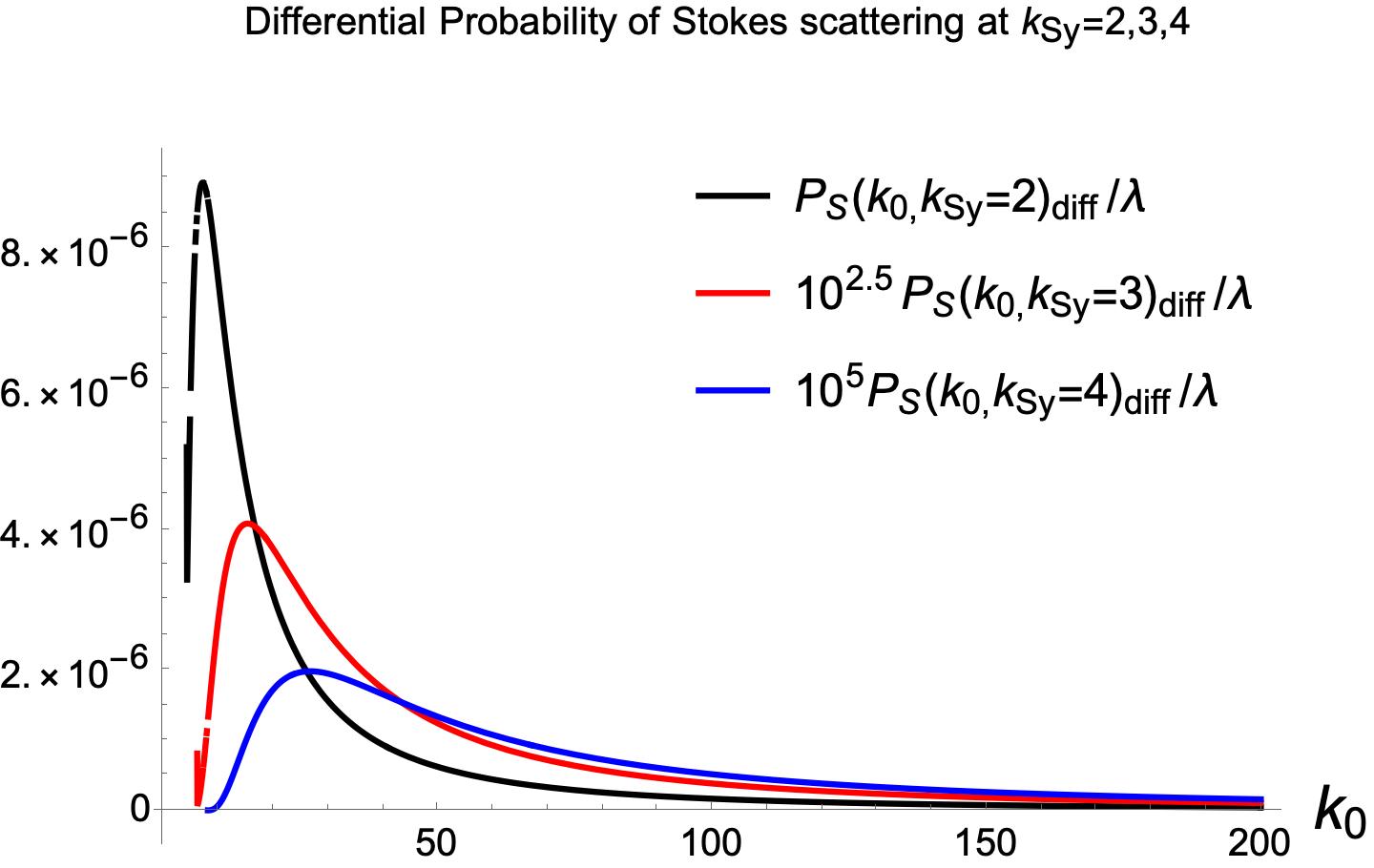}
\includegraphics[width = 0.45\textwidth]{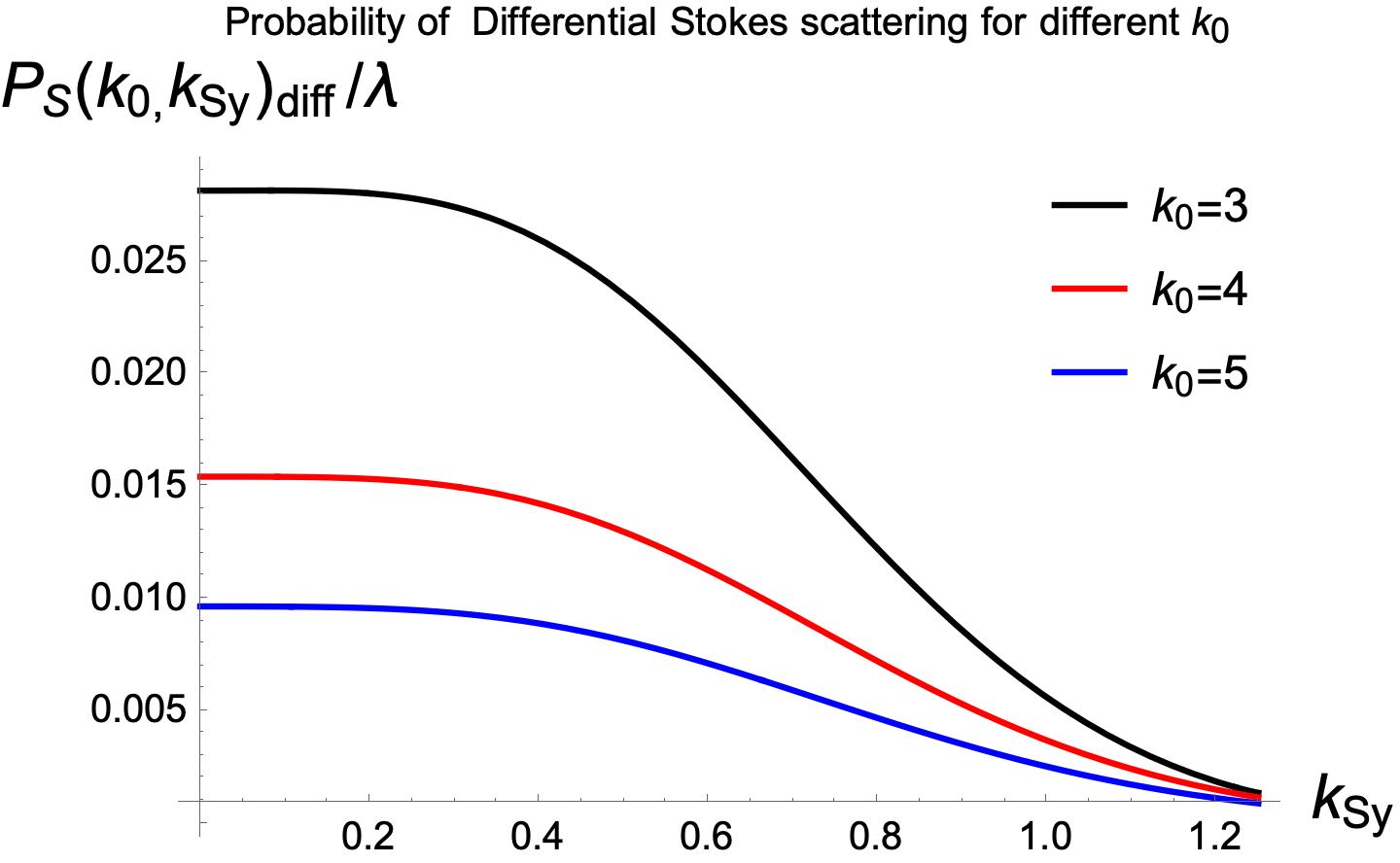}
\caption{The forward, backward  and total probability densities $P_{S}(k_0,k_{Sy})_{\rm{diff}}$ of Stokes scattering with $m=1$. In the left (right) panel, $k_{Sy}$ ($k_0$) is fixed.}\label{2dps2}
\end{figure}

\begin{figure}[htbp]
\centering
\includegraphics[width = 0.9\textwidth]{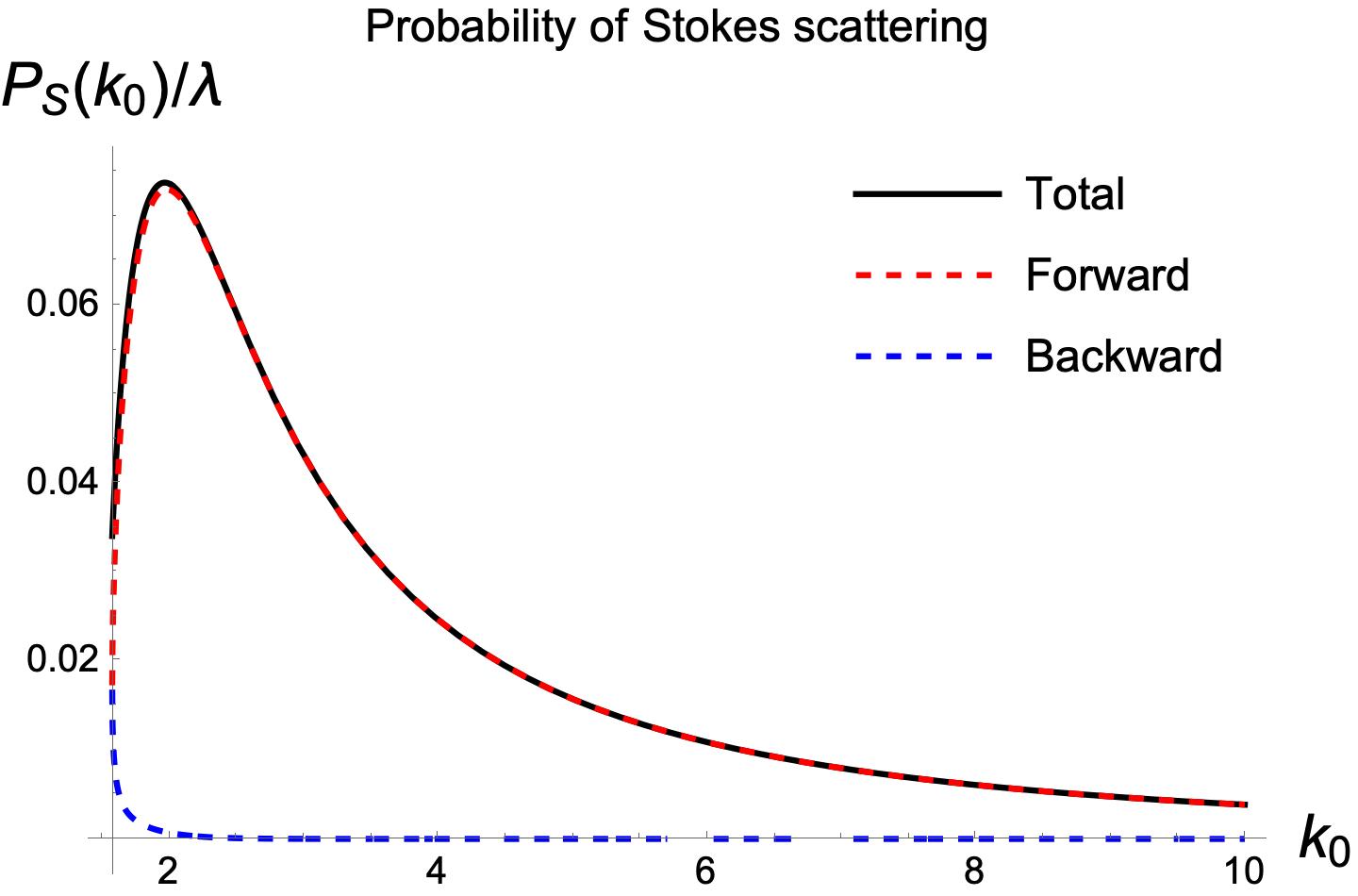}
\caption{The forward, backward and total probabilities $P_{S}(k_0)$ of Stokes scattering with $m=1$.}
\label{2dps3}
\end{figure}

The probability densities for Stokes scattering on a ground-state domain wall are plotted in Fig.~\ref{2dps1} at $k_{Sy}=0$.  For other choices of $k_{Sy}$ and $k_0$, the probability densities are plotted in Fig.~\ref{2dps2}. The total probability of Stokes scattering is plotted in Fig.~\ref{2dps3}.  

\begin{figure}[htbp]
\centering
\includegraphics[width = 0.45\textwidth]{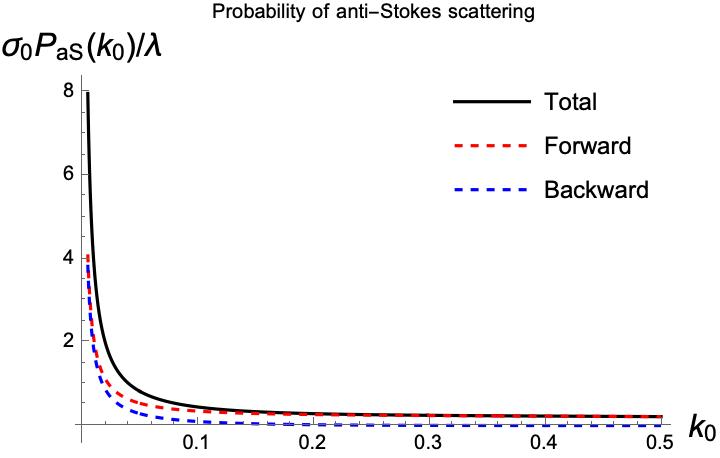}
\includegraphics[width = 0.45\textwidth]{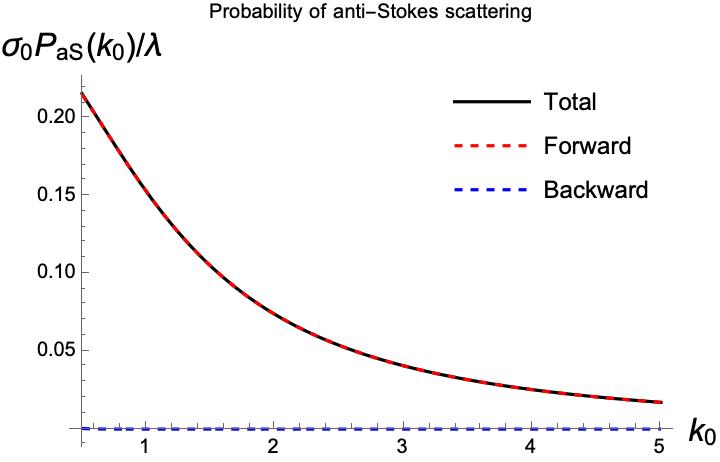}
\caption{The forward, backward  and total probabilities $P_{aS}(k_0)$ of anti-Stokes scattering, with $m=1$. }\label{2dpas}
\end{figure}

The total anti-Stokes scattering probabilities on an excited domain wall are plotted in Fig.~\ref{2dpas}.  These last probabilities are inversely proportional to the size of the initial shape mode wave packet.  To plot them, we multiply the probability by the wave packet size.  The result can be interpreted, up to a factor of order unity, as the cross section of the shape mode for anti-Stokes scattering.  However, to fix this factor, one would have to integrate over distinct impact parameters.  We leave such a study to future work.

We would like to make three observations concerning these numerical results: 

1: Directly from the Figs. \ref{2dps1} and \ref{2dpas},  we see that Stokes scattering is only energetically allowed for a sufficiently high initial momentum $k_0$. Near the threshold, the probability is highest, then it decreases rapidly.  On the contrary, there is no threshold for anti-Stokes scattering. Nonetheless, in the limit of a monochromatic incoming wavepacket, it diverges at small incoming momentum, and then quickly decreases with the incoming meson momentum as is the case for Stokes scattering. 

This behavior is quite similar to that observed for the (1+1)-dimensional kink in Ref.~\cite{Evslin:2023ypw}.  The steep behavior of the probability near the threshold apparently results from the fact that the phase space factor is not differentiable at threshold, as a small increase in $p_0$ leads to a large increase in the maximum allowed outgoing $y$-momenta.

2: Comparing Fig.~\ref{2dps1} with Fig.~\ref{2dps2}, we see the mass threshold for each value of $|k_{Sy}|$ ($ |k_{Sy}|\leq \sqrt{k_0^2+\frac{5}{4}-2\sqrt{1+k_0^2}}$). The differential density of Stokes scattering decreases exponentially as $k_0$ increases at large $k_0$.  This is consistent  with physical principle that, for fixed incoming meson momentum, more energy is available for the y component of the shape mode.  The exponential suppression results from the fact that $x$ momentum needs to be transferred to the domain wall, which is an exponentially suppressed process.  

If the incoming momentum is fixed and as $k_{Sy}$ increases, the probability also decreases because again energy conservation requires a larger $x$-momentum transfer between the meson and the domain wall. 

3: In Fig.~\ref{2dps3}, we see that as the incoming momentum increases above the Stokes scattering threshold, the total probability increases, reaches its maximum and then decreases as incoming meson momentum increases.  Just above the threshold, the probability of forward scattering increases while that of backward scattering decreases and becomes negligible already at $k_0=2m$, which is not far above the threshold.  This is different from the 1+1 dimensional case, in which even the forward probability was maximized at the threshold.  This is to be expected, because in 2+1 dimensions, above the threshold, more phase space is available for Stokes scattering as a result of the $k_y$ degree of freedom.  


\section{Remarks}

When a meson strikes a domain wall string, the leading order interactions occur at order $O(\lambda)$.  There are three, either an additional meson is produced, a translation mode is excited, or a shape mode is excited. The first process is called meson multiplication.  The second has an infrared divergence when the translation mode's longitudinal momentum is small.  In this paper, we have calculated the probability of the last process, Stokes scattering, as well as its inverse process, anti-Stokes scattering, in which the meson de-excites a shape mode that was initially excited.  


We found that both the Stokes and anti-Stokes scattering are very large near their low energy thresholds, but fall rapidly at higher energies.  We expect that, as in 1+1 dimensions, meson multiplication will dominate at high energies.  Indeed, it may grow even faster as a result of the additional phase space afforded by the $y$-momenta. 

We have seen that the dimensionality has limited effect on (anti)-Stokes scattering.
The subleading quantum corrections found for 1+1 dimensional kink case in Ref.~\cite{menorm} that result from the quotient of translation invariance only affect amplitudes at subleading order.  The translation invariance is still not broken in 2+1 dimensions, but nonetheless the behavior of the translation modes is quite different from the (1+1)-dimensional.  As a result, it would be useful to extend the results of Ref.~\cite{menorm} to this setting.

In Ref.~\cite{Evslin:2023ypw}, we studied (anti)-Stokes scattering in the case of kinks in 1+1 dimensions.  In that case the translation mode is a zero mode and the model is gapped.  The (2+1)-dimensional case is more complicated because the translation mode carries a momentum $k_y$ and there is no gap.  In 3+1 dimensions it would correspond to a Goldstone boson and it would break the $x$-translation symmetry.  However in 2+1 dimensions this is forbidden by the theorem of Ref.~\cite{colemanth}.  Physically, in 2+1 dimensions the string fluctuates wildly at large distances in the $y$ direction when an arbitrarily small energy is inserted, potentially ruining our perturbative approach.  

Just how large does $|y|$ need to be before we have problems?  One can see this in our Stokes scattering probability.  Our derivation of the Stokes scattering probability can be applied to the creation of translation modes by simply replacing with $\omega_S$ with $\omega_B=0$.  With this substitution, the probability densities for exciting a translation mode with momentum $k_y$ scales as $\lambda/(m|k_y|)$.  This is no longer small already for $k_y\sim \lambda$, corresponding to the length scale $1/\lambda$ of many processes.  However it is the total probability, not the probability density, which must be small for our perturbative approximation to apply.  If we impose an infrared cutoff  of $\epsilon$ on $|k_y|$, then the total probability is of order $(\lambda/m)$ln$(\epsilon/m)$ which is small so long as $\epsilon>> me^{-m/\lambda}$ corresponding to distances smaller than $(1/m)e^{m/\lambda}$.  This long distance cutoff is quite large as compared with those of any perturbative processes, and so we believe that it does not pose a problem to our semiclassical treatment in this note.

\appendix
\section{Decomposition of the Field} \label{app}
 We expand the Schrodinger picture field and its conjugate momentum in terms of plane waves as

\bea
\phi(\vx)&=&\pinvk\phi_{\vp}e^{-i\vp\vx}\hsp \phi_{\vp}=(\avpd+\frac{\avpm}{\omvp})\nonumber\\
\pi(\vx)&=&\pinvp \pi_{\vp}e^{-i\vp\vx}\hsp \pi_{\vp}=i(\omvp\avpd-\frac{\avpm}{2})
\eea
or, in terms of the domain wall's normal modes $\g_{\vk}(\vx)=\g_{k_x}(\vx) e^{-ik_y y}$ as
\bea
\phi(\vx)&=&\sinvk\phi_{\vk}\g_{\vk}(\vx)\hsp \sinvk=\sxink\pyink\hsp \phi_{\vk}=(\bdvk+\frac{\bvkm}{\omvk})\nonumber\\
\pi(\vx)&=&\sinvk\pi_{\vk}\g_{\vk}(\vx)\hsp \pi_{\vk}=i(\omvk\bdvk-\frac{\bvkm}{2}).\label{decom}
\eea
This is readily invertible
 \bea
 \phi_{\vk}&=&\pinvp\tilde{g}_{-\vk}(\vp)\phi_{\vp}=\pxinp\tilde{g}_{-k_x}(p_x)\phi_{p_x(-k_y)}\nonumber\\
 \pi_{\vk}&=&\pinvp\tilde{g}_{-\vk}(\vx)\pi_{\vp}=\pxinp\tilde{g}_{-k_x}(p_x)\pi_{p_x(-k_y)}
 \eea
 yielding
 \bea
 \phi_{\vp}&=&\sxink\pyink\tilde{g}_{k_x}(p_x)(2\pi)\delta(p_y-k_y)\phi_{\vk}=\sxink\tilde{g}_{k_x}(p_x)\phi_{k_xp_y}\nonumber\\
 \pi_{\vp}&=&\sxink\pyink\tilde{g}_{k_x}(p_x)(2\pi)\delta(p_y-k_y)\pi_{\vk}=\sxink\tilde{g}_{k_x}(p_x)\phi_{k_xp_y}
 \eea
  where $ \tilde{g}_{\vk}(\vx)$ is the Fourier transformation of $\g_{\vk}(\vx)=\g_{k_x}(x)e^{-ik_y y}$
\beq
\tilde{g}_{\vk}(\vp)=\int d^2\vx\gvk(\vx)e^{i\vp\vx}=(2\pi)\delta(p_y-k_y)\tilde{g}_{k_x}(p_x)\hsp \tilde{g}_{k_x}(p_x)=\int dx\g_{k_x}(x)e^{ik_xx}.
\eeq
We then can expand the leading interaction $H\p_3$
\beq
\begin{aligned}
&\int d^2\vx V^{(3)}[gf(x)]\phi(\vx)^3\\
=&\int d^2\vx V^{(3)}[gf(x)]\nsinvk1\nsinvk2\nsinvk3\g_{k_{1x}}\g_{k_{2x}}\g_{k_{3x}}e^{-i(k_{1y}+k_{2y}+k_{3y})y}\phi_{\vk_1}\phi_{\vk_2}\phi_{\vk_3}\nonumber\\
=&\int dxV^{(3)}[gf(x)]\nsxink1\nsinvk2\nsinvk3 \g_{k_{1x}}\g_{k_{2x}}\g_{k_{3x}}\phi_{k_{1x}(-k_{2y}-k_{3y})}\phi_{k_{2x}k_{2y}}\phi_{k_{3x}k_{3y}}.
\end{aligned}
\eeq
Using Eq.~(\ref{decom}), we know the component with $B^{\ddagger}_{Sk_{Sy}}\nbdvk2\nbvk1$ has the form
\beq
\begin{aligned}
\int dxV^{(3)}[gf(x)]\g_{S}(x)\g_{k_{2x}}(x)\g_{-k_{1x}}(x)
\frac{3B^{\ddagger}_{Sk_{Sy}}\nbdvk2\nbvk1}{2\nomvk1}2\pi\delta(k_{2y}+k_{Sy}-k_{1y})\label{a6}
\end{aligned}.
\eeq
 and the component with $\nbdvk2B_{Sk_{Sy}}\nbvk1$ has the form
 \beq
\begin{aligned}
\int dxV^{(3)}[gf(x)]\g_{k_{2x}}(x)\g_{S}(x)\g_{-k_{1x}}(x)
\frac{3\nbdvk2B_{Sk_{Sy}}\nbvk1}{4\nomvk1\omega_{Sk_{Sy}}}2\pi\delta(k_{2y}-k_{Sy}-k_{1y}).\label{a7}
\end{aligned}
\eeq
\section* {Acknowledgement}

\noindent
JE is supported by NSFC MianShang grants 11875296 and 11675223.   HL is supported by the Ministry of Education, Science, Culture and Sport of the Republic of Armenia under the Postdoc-Armenia Program, grant number 24PostDoc/2‐1C009. JE and HL are supported by the Ministry of Education, Science, Culture and Sport of the Republic of Armenia under the Remote Laboratory Program, grant number 24RL-1C047.

\end{document}